\documentclass[fleqn,usenatbib]{mnras}

\usepackage{newtxtext,newtxmath}

\usepackage[T1]{fontenc}
\usepackage{ae,aecompl}

\usepackage{graphicx}	
\usepackage{amsmath}	
\usepackage{amssymb}	
\usepackage{times}
\usepackage{url}
\usepackage{color,graphicx}

\title[The orbital clusters among the { near-Earth} asteroids]{The orbital clusters  among the { near-Earth} asteroids}

\author[T.J. Jopek]{
Tadeusz J. Jopek\thanks{E-mail: jopek@amu.edu.pl}
\\
Astronomical Observatory Institute, Faculty of Physics, A. Mickiewicz University, 
Sloneczna 36, 60-286 Pozna\'{n}, Poland
}

\date{Accepted 2020 March 9 . Received 2020 March 8; in original form 2019 November 26}

\pubyear{2019}

\definecolor{ziel}{rgb}{0.93,0.90,0.50}
\newcommand{\sz}[1]{\bf {#1}}

\begin{document}
\label{firstpage}
\pagerange{\pageref{firstpage}--\pageref{lastpage}}
\maketitle

\begin{abstract}
 Fifteen orbital clusters (associations) were identified  among $\sim$$20000$ { near-Earth} asteroids (NEAs). 
 All associations were found with a high statistical reliability { using} { a} single linkage cluster analysis algorithm and three orbital similarity functions. 
 { The identified} groups { constitute a}  small fraction ($4.74$\%) of the { entire} sample. { Notwithstanding, they may be} hazardous to { Earth} and its inhabitants. 
{ As with meteoroid streams, every year Earth} comes very close to the orbits of each association. In two cases (2008TC3 and 2017FU102), the distance between the { asteroid's orbit and Earth's orbit} was { shorter than Earth's radius.} Among { the members} of the identified associations, we found $331$ objects larger than the Chelyabinsk asteroid and all {of them} approach { Earth's} orbit at a distance smaller than $0.05$ [au]. 
 Two of the identified groups, (4179) Toutatis and (251430) Itokawa, { support} {the}
 { hypothesis regarding the catastrophic origins of the asteroids Toutatis and Itokawa through violent collisions.}
 
This study does not focus on the origin of the NEA associations, {  but rather focuses on tracing the associations}. { Regardless} of their origin, the identified groups pose a serious threat to {  Earth}. Hence, to facilitate their monitoring of we { calculated} coordinates of the theoretical radiants and { the} calendar { dates} of their potential activity.
\end{abstract}




\section{Introduction}
\label{intro}
{ While} the existence of { main-belt} asteroid (MBA) families is accepted beyond doubt, { the} existence of families among { near-Earth} asteroids (NEAs) is an open issue, { as is} the problem of the genetic association of comet groups and pairs. 
In \citet{1982BAICz..33..150K} { the} author wrote ``{\it ... to illustrate the diversity of opinions on this issue, it may be quoted  from  \citet{1971IrAJ...10...35O}, that there are at least $60$ groups of two or seven known comets which are due to real genetic associations.'' ``... and from \citet{1977Icar...30..736W}, that except for a few pairs these groups exhibit similarity in their orbital elements that is no greater than might 
 be expected by chance.}''

{ The search} for { groupings} { among} the NEAs { does not have a} long history. Almost twenty years have passed since an extensive search { was performed}  by \cite{2000Icar..146..453D}, { who found 14 associations of 4-25 members among 708 NEAs, more than half of them might
be attributed to chance alignments.}
This study was followed { by that of}  \cite{2005Icar..178..434F}, { who} made { a} critical study of Drummond's results and confirmed his scepticism { in noting that}  ''... Drummond's families are nothing more than random fluctuations in the distributions of NEA osculating orbital elements``. \cite{2012Icar..220.1050S} { in} searching for families { among} $\sim$$7500$ { near-Earth} objects (NEOs), { confirmed} { that} they had not identified any NEO { families}. 
{ Different} results were { obtained} by \cite{2011epsc.conf...15J, 2015HiA....16..474J} and recently by \cite{2016MNRAS.456.2946D}. Jopek { found} more than ten groups of more than ten members { among $\sim$$9000$ } NEAs. \cite{2016MNRAS.456.2946D} { examined} the orbits of $\sim$$13500$ NEOs { and as a result} { they} confirmed the presence of statistically significant dynamical groupings among the NEO population.

The { aforementioned} studies { produced} different results, which is understandable as { the} search for families (members originating from the same parent body) and { the} search for associations (members displaying { only} the orbital similarity) are two different tasks.      
{ Notwithstanding,} both { the} ample families and { the} associations pose an increased threat to { Earth} regardless of their origin, { hence} the search for grouping  { among} the NEA's is an important issue.

In this study, { an extensive and statistically strict search for associations has been conducted} { among} $\sim$$20000$ NEA orbits by applying the cluster analysis method, { which was} developed and validated for { searching} the meteoroid stream \citep{2011MmSAI..82..310J}.
\section{NEA data and the searching method}
\subsection{NEA orbital sample}
The orbital data { of the NEAs} were retrieved { on} May 14th, 2019 from the NEODyS-2 database.\footnote{The database is publicly available at 
the URL \url{https://newton.spacedys.com/neodys/index.php?pc=5}}
{ Heliocentric} { ecliptic} osculating elements (epoch 58400 MJD [TDB], reference frame J2000): $a$~---~semi-major axis, $e$ --- eccentricity, $q$ --- perihelion distance, $\omega$ --- argument of perihelion, $\Omega$ --- longitude of ascending node and $i$ --- inclination {were used}. The orbits with $i > 90^\circ$ or $a > 4.4$ [au] or  $q > 1.35$ [au]  ---  $31$ { in total} were  discarded. Additionally, two objects (2010 LB67, 2010 LB64) { whose} absolute magnitude $H$$=$$-1$ { were rejected}. Accordingly,  in this study $20038$ osculating orbits { were} searched for { groupings}.

\begin{figure}
\centerline{
\includegraphics[width=0.32\textwidth]{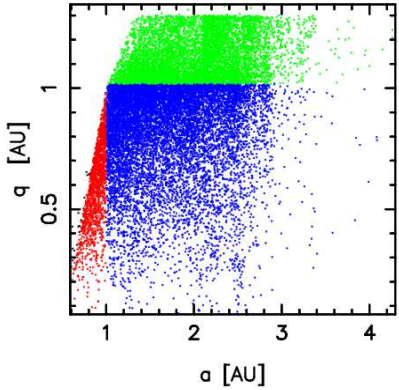}
}
\caption{Distribution of $20038$ NEAs { on} { an} $a$-$q$ plane. $7471$ green points represent the Amor sub-sample, $11032$ blue points represent the objects of the Apollo type, $1516$ red points represent the Aten sub-sample. $19$ black points (just above the Atens) mark the Atira objects.
(For interpretation of colours in this figure the reader is referred to the web version of this article.)
}
\label{fig:all-a-q}
\end{figure}

\begin{figure}
\begin{center}
\vbox{
\includegraphics[width=0.30\textwidth]{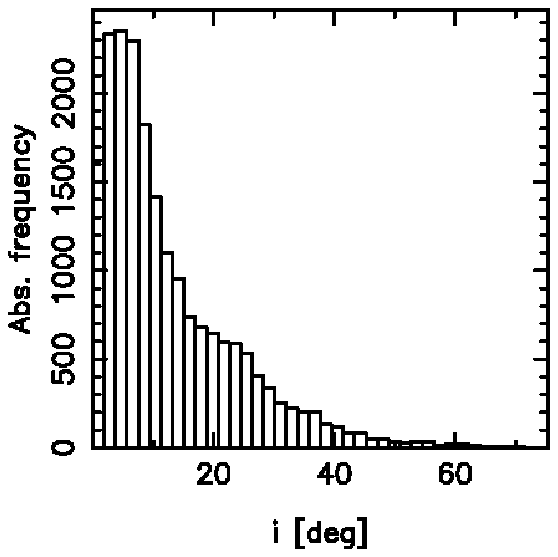}     
\includegraphics[width=0.315\textwidth]{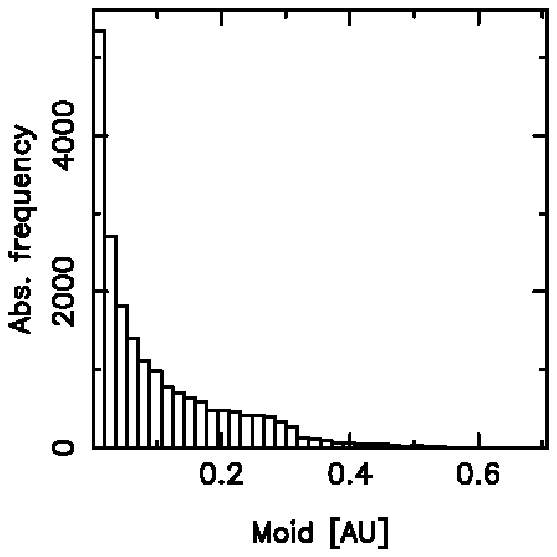}
}
\end{center}
\caption{Histograms of { the} orbital inclinations and the MOIDs of $20038$ NEAs. { Low-inclined} orbits predominate,  the median $i_{0.5}=9^\circ$. { Small} MOIDs { are  also predominant}, the median MOID$_{0.5}=0.0550$ [au].}
\label{fig:hist_im}
\end{figure}

The { sample of} NEAs contains four types of objects (see Fig \ref{fig:all-a-q}), namely: $19$ Atiras, $1516$ Atens, $11032$ Apollos and $7471$ Amors, as defined by separate regions on the  $a$-$q$ plane they occupy\footnote{We applied definitions given at \url{https://cneos.jpl.nasa.gov/about/neo_groups.html}}.  
{ The sample of} NEAs is dominated by { low-inclination} orbits (see Fig.~\ref{fig:hist_im}), { with} more than half { having} { an angle of} inclination { which is} smaller than $10^\circ$. The median $i_{0.5}=9^\circ$, the maximum value equals $i_M$=$75^\circ$.

For each NEA orbit, we calculated the minimum orbital intersection distance (MOID) with { Earth's} orbit, along with corresponding theoretical geocentric radiant parameters. The MOIDs  and the radiant coordinates were calculated { according to} the methods { provided}  in \citet{1986CeMec..38..345D} and \citet{1964AcA....14...25S}, respectively. { Earth's} orbital elements were calculated on the osculating epoch $2458400.5$ [TDB], using $DE405$ JPL Ephemerides\footnote{We used the software and { the} data file from \url{https://ssd.jpl.nasa.gov/}
}. 

In the { sample of} NEAs small MOIDs { are predominant} --- the median is MOID$_{0.5}=0.0550$ [au], which is close to the value given by the NASA definitions { of Potentially} Hazardous Asteroids (PHA). Applying both { criteria} for PHAs: MOID $\leq0.05$ [au] and absolute magnitude $H\leq22.0^m$ we found $1969$ PHAs, which { constitutes} $9.8\%$ of the whole sample. 
Assuming { the} albedo of $0.15$, the value $H=22^m$ corresponds to the asteroid's diameter $d=140$ [m]. However, if we choose $H=26^m$ (the diameter $d=22$ [m]), the number of NEAs satisfying the conditions $H\leq26^m$ and MOID $=0.05$ [au] rises to $6825$ objects, $34\%$ of the { sample of} NEAs.  It is { significant}  to note that { the} Chelyabinsk { asteroidal impactor has a} diameter $\sim$$20$ [m] \citep{2013Sci...342.1069P}.
\subsection{Searching method}
Our { search} method comprises of three components:
\vspace{-0.2cm}
\begin{itemize}
\item {\em cluster analysis algorithm} --- we used a single linkage method (a { variant} of { the} general hierarchical cluster analysis method) proposed for the meteoroid stream identification by \citet{1963SCoA....7..261S}.
 \item {\em orbital similarity function} --- we used the hybrid D-function $D_H$ proposed by \citet{1993Icar..106..603J} and two $\rho$--functions ($\rho_1, \rho_2$), applied in  \citet{2016MNRAS.462.2275K}. 
 { The} mathematical { formulae} and { further} details about different $D$- and $\rho$- functions { can be found} in \citet{2019msme.book..210W}.
 \item {\em orbital similarity thresholds} were determined using a statistical approach; the thresholds were found separately for each group of $2,3, ...,M$=$50$ members, and for each { of the} $D_H,\rho_1, \rho_2$ functions. All thresholds corresponded to { a} low probability (less than $1$\%) of chance grouping.   
\end{itemize}
The orbital similarity thresholds were determined by {using a} numerical experiment (similar to { that which is} described by \citet{1997A&A...320..631J, 2003MNRAS.344..665J}), namely:
\begin{itemize} 
 \item we searched for { groupings} in the synthetic orbital samples starting from threshold $D_0$ for which no single group of $2,3,4 ...$ members { was} found, 
 \item next, { the search} for groups was repeated with { an} increased value of $D_0$ until the first { groups} of $2,3,4,..., 50$ members, respectively, were found and { the} obtained $D_{0M}$ corresponding to the previous step were stored, 
 \item when the last $D_{0M}$ value was found the new synthetic orbital sample was searched. For each distance function $100$ synthetic orbital samples were searched.
 \item The final individual thresholds $D_{M}$, $M=2...50$ were calculated as the arithmetic means of one hundred of { the} $D_{0M}$ values. 
\end{itemize}

The synthetic NEA orbits were generated using { the} inversion of the cumulative probability distributions of the observed sample, { according to} the { C} method { which is} described in \citet{2017P&SS..143...43J}. Only { the} synthetic orbital samples  which had passed the consistency test with the observed sample { were used}. The { chi-squared} test was applied, see \citet{2002numrec.book.....P}.
Before the final thresholds $D_{M}$ were determined, the possible groups in the observed { sample of NEAs} { had been}  excluded. They { had been} identified by { using} the same cluster analysis as described above. After { the} elimination, the orbital distributions of the observed reduced sample were used to determine the final  thresholds $D_{M}$.

\begin{figure}
\centerline{
\includegraphics[width=0.5\textwidth]{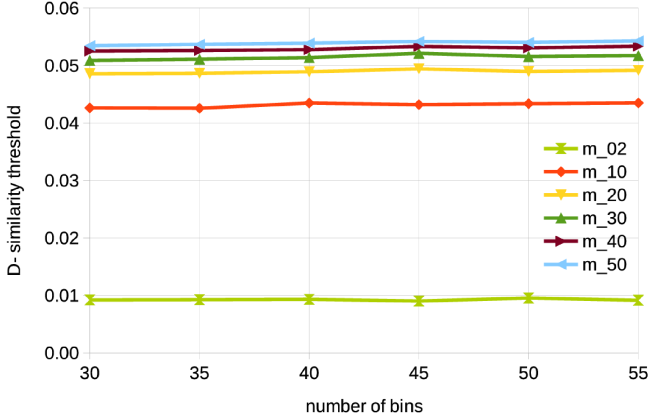}
}
\caption{The orbital similarity thresholds determined for { the} $\rho_1$ function versus different { histogram bin} sizes used for { the} synthetic { orbit} generation. The thresholds are plotted for the groups of $2,10,20,30,40,50$ members identified among the synthetic NEA orbits. }
\label{fig:wplywlpk}
\end{figure}

To estimate the sensitivity of $D_{M}$ values to the assumed parameters of our synthetic { orbit-generation} method, we { carried out} several additional simulations. We checked the influence of the size of the histogram bins and the choice of the seed of the uniform random number generator.  Fig.~\ref{fig:wplywlpk} shows that the thresholds (an example for the $\rho_1$ function is given) are not sensitive to the { histogram bin size}. A similar { lack of} dependence was found { while} changing the seeds of the random number generator, { as well as} the remaining functions. 

{ Finally,} we determined the medians of the $D_{M}$ values { that were} obtained from different bins and seeds. Their arithmetic means are { provided} in Table \ref{tab:thresholds}, and these { are the} values { that are} applied in this study.
\begin{table}
\caption{The values of the orbital similarity thresholds and their uncertainties applied in the { search} for associations { among} $20038$ NEA orbits. { The} thresholds correspond to the reliability level $>99\%$ { for}  each group of $M$ members and for each distance function: $D_{H}, \rho_1, \rho_2$ . The thresholds { provided} in this table are closely related with the { single-linkage} method and the size of the NEA sample used in this study. Before being { used,} the thresholds { had been} multiplied by the factors given in the second row of the table.}
\begin{center}
\scriptsize
\begin{tabular}{lc@{}cc@{}cc@{}cc@{}cc@{}cc@{}ccc@{}c}
\hline
\multicolumn{1}{l}{} &  \multicolumn{2}{l} {$D_{H} \pm \sigma$} 
                     &  \multicolumn{2}{l} {$\rho_{1} \pm \sigma$} 
                     &  \multicolumn{2}{l} {$\rho_{2} \pm \sigma$}\\
\hline
\multicolumn{1}{l}{$M$}   & \multicolumn{2}{l}{\,\,$10^{-2}$\,\,\,\,\,\,\,$10^{-4}$}
                          & \multicolumn{2}{l}{\,\,$10^{-2}$\,\,\,\,\,\,\,$10^{-4}$}
                          & \multicolumn{2}{l}{\,\,$10^{-2}$\,\,\,\,\,\,\,$10^{-4}$}  \\
\hline		
2	&	0,8209	&	\,\,\,	1,0	&	0,9366	&	\,\,\,	2,4	&	0,9951	&	\,\,\,	1.8	\\
3	&	2,0137	&	\,\,\,	4,3	&	2,2837	&	\,\,\,	3,2	&	2,4106	&	\,\,\,	4.1	\\
4	&	2,6528	&	\,\,\,	3,1	&	2,9953	&	\,\,\,	4,1	&	3,1673	&	\,\,\,	4.1	\\
5	&	3,0358	&	\,\,\,	2,7	&	3,4259	&	\,\,\,	4,5	&	3,6307	&	\,\,\,	3.0	\\
6	&	3,2888	&	\,\,\,	1,9	&	3,7021	&	\,\,\,	3,7	&	3,9323	&	\,\,\,	3.2	\\
7	&	3,4668	&	\,\,\,	1,2	&	3,9008	&	\,\,\,	3,1	&	4,1425	&	\,\,\,	2.2	\\
8	&	3,6065	&	\,\,\,	1,5	&	4,0659	&	\,\,\,	2,6	&	4,3299	&	\,\,\,	2.3	\\
9	&	3,7268	&	\,\,\,	2,5	&	4,2044	&	\,\,\,	3,7	&	4,4688	&	\,\,\,	2.4	\\
10	&	3,8240	&	\,\,\,	3,1	&	4,3102	&	\,\,\,	2,7	&	4,5966	&	\,\,\,	1.6	\\
11	&	3,9039	&	\,\,\,	2,3	&	4,4022	&	\,\,\,	3,4	&	4,6759	&	\,\,\,	1.4	\\
12	&	3,9739	&	\,\,\,	2,3	&	4,4809	&	\,\,\,	3,0	&	4,7628	&	\,\,\,	1.8	\\
13	&	4,0355	&	\,\,\,	1,6	&	4,5455	&	\,\,\,	2,4	&	4,8336	&	\,\,\,	2.4	\\
14	&	4,0870	&	\,\,\,	1,4	&	4,6078	&	\,\,\,	2,2	&	4,8916	&	\,\,\,	2.2	\\
15	&	4,1400	&	\,\,\,	1,9	&	4,6584	&	\,\,\,	2,9	&	4,9490	&	\,\,\,	2.3	\\
16	&	4,1807	&	\,\,\,	1,8	&	4,7049	&	\,\,\,	3,2	&	5,0058	&	\,\,\,	3.0	\\
17	&	4,2194	&	\,\,\,	2,0	&	4,7521	&	\,\,\,	2,5	&	5,0524	&	\,\,\,	2.7	\\
18	&	4,2522	&	\,\,\,	2,3	&	4,7889	&	\,\,\,	2,5	&	5,0958	&	\,\,\,	2.1	\\
19	&	4,2904	&	\,\,\,	1,6	&	4,8248	&	\,\,\,	2,3	&	5,1369	&	\,\,\,	1.6	\\
20	&	4,3224	&	\,\,\,	2,0	&	4,8560	&	\,\,\,	2,7	&	5,1747	&	\,\,\,	1.7	\\
21	&	4,3466	&	\,\,\,	1,9	&	4,8859	&	\,\,\,	2,7	&	5,2068	&	\,\,\,	2.1	\\
22	&	4,3712	&	\,\,\,	1,9	&	4,9133	&	\,\,\,	2,2	&	5,2359	&	\,\,\,	2.6	\\
23	&	4,4000	&	\,\,\,	2,2	&	4,9453	&	\,\,\,	2,6	&	5,2659	&	\,\,\,	2.6	\\
34	&	4,4229	&	\,\,\,	2,5	&	4,9706	&	\,\,\,	2,8	&	5,2927	&	\,\,\,	2.7	\\
25	&	4,4425	&	\,\,\,	2,5	&	4,9915	&	\,\,\,	2,8	&	5,3193	&	\,\,\,	3.0	\\
26	&	4,4664	&	\,\,\,	2,7	&	5,0131	&	\,\,\,	2,8	&	5,3452	&	\,\,\,	2.8	\\
27	&	4,4845	&	\,\,\,	2,7	&	5,0336	&	\,\,\,	3,1	&	5,3692	&	\,\,\,	3.0	\\
28	&	4,5019	&	\,\,\,	2,5	&	5,0492	&	\,\,\,	2,9	&	5,3889	&	\,\,\,	3.1	\\
29	&	4,5213	&	\,\,\,	2,8	&	5,0669	&	\,\,\,	2,9	&	5,4103	&	\,\,\,	3.2	\\
30	&	4,5382	&	\,\,\,	2,9	&	5,0853	&	\,\,\,	2,8	&	5,4300	&	\,\,\,	3.5	\\
31	&	4,5518	&	\,\,\,	2,6	&	5,1009	&	\,\,\,	2,6	&	5,4481	&	\,\,\,	3.0	\\
32	&	4,5631	&	\,\,\,	2,5	&	5,1157	&	\,\,\,	2,5	&	5,4614	&	\,\,\,	2.9	\\
33	&	4,5760	&	\,\,\,	2,5	&	5,1308	&	\,\,\,	2,5	&	5,4791	&	\,\,\,	2.6	\\
34	&	4,5908	&	\,\,\,	2,6	&	5,1454	&	\,\,\,	2,8	&	5,4934	&	\,\,\,	2.6	\\
35	&	4,6009	&	\,\,\,	2,7	&	5,1583	&	\,\,\,	2,7	&	5,5086	&	\,\,\,	2.6	\\
36	&	4,6147	&	\,\,\,	3,0	&	5,1713	&	\,\,\,	2,6	&	5,5245	&	\,\,\,	2.7	\\
37	&	4,6288	&	\,\,\,	3,1	&	5,1856	&	\,\,\,	2,3	&	5,5388	&	\,\,\,	2.5	\\
38	&	4,6424	&	\,\,\,	3,1	&	5,1966	&	\,\,\,	2,3	&	5,5521	&	\,\,\,	2.7	\\
39	&	4,6541	&	\,\,\,	3,0	&	5,2113	&	\,\,\,	2,3	&	5,5678	&	\,\,\,	2.9	\\
40	&	4,6643	&	\,\,\,	2,9	&	5,2233	&	\,\,\,	2,2	&	5,5777	&	\,\,\,	2.8	\\
41	&	4,6744	&	\,\,\,	3,1	&	5,2360	&	\,\,\,	2,0	&	5,5879	&	\,\,\,	2.7	\\
42	&	4,6838	&	\,\,\,	3,1	&	5,2475	&	\,\,\,	2,2	&	5,6010	&	\,\,\,	2.8	\\
43	&	4,6917	&	\,\,\,	3,1	&	5,2562	&	\,\,\,	2,0	&	5,6145	&	\,\,\,	2.7	\\
44	&	4,7025	&	\,\,\,	3,0	&	5,2664	&	\,\,\,	1,9	&	5,6273	&	\,\,\,	2.5	\\
45	&	4,7127	&	\,\,\,	2,7	&	5,2785	&	\,\,\,	1,9	&	5,6383	&	\,\,\,	2.4	\\
46	&	4,7215	&	\,\,\,	2,9	&	5,2880	&	\,\,\,	2,1	&	5,6485	&	\,\,\,	2.3	\\
47	&	4,7304	&	\,\,\,	2,9	&	5,2972	&	\,\,\,	2,0	&	5,6566	&	\,\,\,	2.0	\\
48	&	4,7386	&	\,\,\,	3,0	&	5,3066	&	\,\,\,	2,3	&	5,6672	&	\,\,\,	2.1	\\
49	&	4,7454	&	\,\,\,	2,8	&	5,3170	&	\,\,\,	2,2	&	5,6761	&	\,\,\,	2.0	\\
50	&	4,7529	&	\,\,\,	2,7	&	5,3248	&	\,\,\,	2,0	&	5,6851	&	\,\,\,	1.9	\\
\hline
\label{tab:thresholds}
\end{tabular}
\end{center}	
\normalsize
\end{table}
\section{General results}
{ The results of all} nine searches are summarised in  Table~\ref{tab:genresults}. { We found} $18$--$22$ associations, which { constitutes} $3.6$--$5.1$\% orbits of the whole { of the sample of} NEAs. { The obtained}  percentage values are considerably { lower} than in the meteor streams { search}, e.g.: in \citet{1963SCoA....7..261S} of the 359 photographic orbits about $36$\%  { belonged} to streams, in \citet{1986PhDT........61J} of the 1608 photographic orbits about $58$\% { belonged} to streams, in \citet{1993mtpb.conf..269J} of the $531$ TV meteor orbits about $30$\% { belonged} to stream component, in \citet{1999md98.conf..307J} of the $3675$ radio orbits about $28$\% { belonged} to { streams, and} in \citet{2003MNRAS.344..665J} of the 256 bolides about $29$\% { belonged} to streams. Of  $\sim$$19000$ radio-meteoroids only about $16$\% { belonged} to streams \citep{1976Icar...27..265S}.

For each individual distance function, the results are quite consistent; { and } the results obtained with $D_t\pm \sigma$ thresholds are similar. However, { the results} differ when different functions are compared. 
{ Most of the} associations ($22$ of them) were found using the $\rho_1$ function. { Their} sizes were rather small { and} the most numerous group { consisted} of $292$ members. { Moreover, when} using { the} $\rho_1$ function { the} associations made { up} the highest $5.1$\% of the total sample.
Concerning the number of identified groups the searches with { the} $D_H,\rho_1, \rho_2$ gave reasonably similar results. { The most} stable results were obtained with { the} $\rho_1$ function { while} the largest groups (above $400$ members) were identified with { the} $\rho_2$ metrics.  

Such diverse results hamper the interpretation of the cluster analysis, but in the case of small bodies of the Solar System such differences are not surprising. Similar differences were observed in our earlier studies described in \citet{2011epsc.conf...15J, 2011MmSAI..82..310J, 2015HiA....16..474J}. They are caused by different properties of the orbital distance functions. Different  D- and $\rho-$ functions are not equivalent to each other \citep[see][]{1993Icar..106..603J, 2016MNRAS.462.2275K}. They are expressed by different mathematical { formulae} and depend on different { numbers} of variables: five { for the} $D_H$ and { the} $\rho_1$, six { for the} $\rho_2$ function. These sets of variables refer to different spaces, therefore it is { not surprising that when} using distinct orbital distance functions we { obtain} different results { from} the cluster analysis. Hence, the problem is { in deciding upon} how { we} should extract { the most robust associations} from this complicated picture.
\begin{table}
\begin{center}
\caption{The overall results of nine cluster { analyses} among $20038$ NEAs. $N_G$ is the total number of identified groups. $N_A$ is the total number of asteroids identified in $N_G$ groups; $P$ is the fraction of the group component in the whole sample. { The} last two columns { provide} the minimum and maximum sizes of the identified groups. For each distance function the results correspond to three thresholds values: $D_{t}$, $D_{t}-\sigma$ and $D_{t}+\sigma$ (see Table~1).}
\begin{tabular}{l c c c rr}
\hline
\multicolumn{1}{c}{Dist. func.} & \multicolumn{1}{c}{$N_G$} & 
\multicolumn{1}{c}{$N_A$}  & \multicolumn{1}{c}{$P [\%]$} & \multicolumn{1}{c}{ $Min$} &\multicolumn{1}{c}{$Max$}    \\
\hline
$D_{H}$	&	21	&	798	&	4.0	&	2	&	211	\\
$D_{H}+\sigma$	&	20      &	885	&	4.4	&	2	&	261	\\
$D_{H}-\sigma$	&	21 	&	714	&	3.6	&	2	&	154	\\
\hline
$\rho_{1}$	        &	22	&	954	&	4.8	&	2	&	283	\\
$\rho_{1}+\sigma$	&	22	&	1025	&	5.1	&	2	&	292	\\
$\rho_{1}-\sigma$	&       22	&	942	&	4.7	&	2	&	281	\\
\hline
$\rho_{2}$      	&	20	&	903	&	4.5	&	2	&	495	\\
$\rho_{2}+\sigma$	&	20	&	970	&	4.8	&	2	&	602	\\
$\rho_{2}-\sigma$	&	18	&	837	&	4.2	&	2	&	495	\\
\hline
\label{tab:genresults}
\end{tabular}
\end{center}
\end{table}

In what follows we discuss the results obtained with { the} $D_{H}$, $\rho_{1}$ and $\rho_{2}$ { functions}.
No single association of $3$ or more members was found by all functions. { Some} groups were found using { only} one distance function, but for many groups a partial { overlap} occurred. { The most} consistent result, { which was} found by all functions, are two very compact pairs: 2015FO33, 2015YZ and 2017SN16, 2018RY7. 
On the other hand, one { very} complicated result is the huge association found { using} { the} $\rho_2$ function with the threshold $D_t$$=$$0.0566$. { It} is designated by { the} $\rho_2/296$ { code} and consists of $495$ members  
(in this study we designate association by the D-function and { a} running number, in our data sample, of the first NEA which belongs to this group).
{ The} $\rho_2/296$ association was partly split into $2$ sub-groups found by { the} $D_H$ { as well as} by { the} $\rho_1$ function. { Moreover}, { the} $\rho_2/296$ association contained { many} NEA's which were not included in the sub-groups in question. The $\rho_2/296$ { proved} to be our { most} complex finding. { We} observed similar splitting { in other large groups}. 

In the next section, we { briefly} present the results obtained by other researchers { before providing more details about our { own} findings}.
%
\subsection{The results obtained before this study}
We refer to { those} studies { where} authors performed extended { searches} for groups { that were}  exclusively among the { orbits of the} NEAs.

In \citet{2000Icar..146..453D} the author { analysed} $708$ orbits { by applying} a different searching method, using { the} single linkage approach only when searching for pairs and strings. { He also} used the $D_{SH}$ function proposed by \citet{1963SCoA....7..261S} and a single value of the orbital similarity threshold $D_t$$=$$0.115$. Drummond found $14$ associations of $4$-$25$ members, $8$ strings (i.e. groups of three members) and $7$ pairs. His { greatest} association, $A1$, consisted of $25$ members. All groups { from the entire sample of NEAs} contained $22$\% { of the} orbits, whereas
in our study this percentage { was only} $3.6$-$5.1$\%. In our study we found many members of Drummond groups, however our findings were different, e.g. in our study Drummond's $A1$ association { was} split up into $DH/178$ ($5$ members) and $\rho_1/659$ ($1$ member).

As mentioned in section \ref{intro}, \citet{2012Icar..220.1050S} searched { for families} among $7500$ NEAs. They applied different cluster { analyses}  and $D_{SH}$ function. Three clusters  $C1, C2, C3$ { were found,} { they consisted} of $4$, $6$ and $5$ members only and no cluster passed the strict significance { threshold for the family test}. In { our} search, none of { the} four objects of $C1$ group { was} identified as { an} association member; five of { of the} six members of $C2$ were in $DH/3058$  and $\rho_1/3058$. All five members of $C3$ were in our $D_H/955$. In our study, however, we do not claim that our groups which incorporate $C2, C3$ are NEA families.

In our earlier search described in \citet{2011epsc.conf...15J, 2015HiA....16..474J, 2019Viena} we used { a} similar approach { to that used in the present} study, however, only very limited and general results have been presented. 
{ At the} XXVIII~IAU~GA in Beijing \citep{2015HiA....16..474J} we presented { the} results of { the} search among $9004$ NEAs. We { accepted} $10$ associations which contained $\sim$$8$\% orbits of the whole sample. {  These findings are referred to} in the following section.

Quite { recently,} \citet{2016MNRAS.456.2946D} searched for groupings among { the orbits of} $\sim$$13400$ NEAs. { Those authors} applied another cluster analysis algorithm as well as two different D-functions, { namely,} the reduced $D_R$ function proposed by \citet{1999MNRAS.304..743V} and { a} simplified form of the $D_{SH}$ function proposed by \citet{1994ASPC...63...62L}. { Five} groups consisting of  $43$-$180$ members { are reported in their paper}. { As} the authors did not { provide} a full list of members of the identified groups (only a few members were specified) it is difficult for us to refer to this study. We were only able to find that { the $4$ mentioned} NEAs { from} their 5011 Ptah group do not belong to any group found in our search; from $3$ NEAs mentioned as the 85585 Mjolnir  group only one was found in our $\rho_1/3041$, $\rho_2/3041$ associations and from $2$ members of the 101955 Bennu group mentioned by \citet{2016MNRAS.456.2946D} only one { was} found in our $\rho_2/296$ association.
\section{Results and discussion}
In this section we describe { the} details of $15$ associations { that are} identified in this study. 
{ The { associations} were named after} the first asteroid in the NeoDys catalogue { belonging} to the identified group. { Our} working  code was { also} assigned --- { it}    represents the distance function used, and { a} running number of the first orbit in our NEA catalogue { that was} assigned to the group. 
For each group { the members' names were listed}. In bold { type} are the names of the NEAs objects, which satisfy the { following} conditions --- the absolute magnitude $H\leq26$ and the MOID$\leq0.05$ [au]. We call such objects { `Chelyabinsk class objects`, which indicates { that} they pose a potential threat to { Earth's} inhabitants { which is} comparable to that of} {the} Chelyabinsk { asteroidal impactor}.
The orbits belonging to the same  association are { illustrated graphically}. { We applied different orbital projections in a { drawn} plane}. The orbital arcs running below the ecliptic { plane} are plotted in blue. All physical parameters { such as} absolute magnitude $H$ and diameters $d$ were taken from \url{http://neo.ssa.esa.int/} or  \url{http://neo.ssa.esa.int/neo-home}.
\subsection{(2061) Anza association}
Anza association (our internal code $D_H/23$) was named { after} Anza NEA of absolute magnitude $H$$=$$16.38^m$ and diameter $d$$=$$1700$ [m].
The group consists of $89$ members (see Table~\ref{tab:Anza}). { It} was identified using { the} $D_H$ function with the orbital similarity threshold $D_t=0.047386$. Anza was found in our earlier search among $\sim$$9000$ NEAs \citep[see][]{2015HiA....16..474J}, at that time only $34$ orbits of the group { had been} identified.

\begin{figure}
\centerline{
\includegraphics[width=0.30\textwidth]{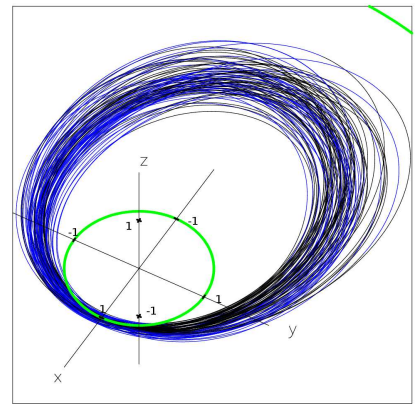}
}
\caption{Anza (2061) association. $89$ orbits resemble a meteoroid stream approaching { Earth's} orbit { at} distances { of} $0.0006$-$0.1063$ [au]. { The} activity of the bolides { that are} potentially related with this group starts in August and ends in November. The orbital arcs running below the ecliptic are plotted in blue. In green, the projection of { Earth's} and { Jupiter's} orbits are plotted. { Earth's} orbit lies inside Anza association. In the bottom right corner the arc of  { Jupiter's} orbit can { be} seen. 
The units on the axes correspond to 1 AU.  
}
\label{fig:Anza23}
\end{figure}

The orbit of 2008VB4 ($H$$=$$28.32^m$), { a} member of this association,  approaches { Earth's} orbit very closely. { Its} MOID equals $0.000562$ [au], or $0.219$ [LD] i.e. the mean Earth-Moon distance. { The} parameters of { the} theoretical radiant of this NEA are: the equatorial coordinates  $\alpha_G$$=$$251^\circ$, $\delta_G$$=$$-22^\circ$ and the geocentric velocity $V_G$$=$$11.9$ [km/s]. Potential meteor-bolide activity related with this radiant falls on November 4th. 
{ The} remaining orbits of the Anza group do not approach { Earth's} orbit so closely. { \bf Notwithstanding,} many of them reach MOIDs smaller than $0.05$ [au]; the maximum MOID is $0.1063$ [au], the average is $0.0326$ [au]. 
$38$ members { of this group fall into} { the} Chelyabinsk class, $5$ objects meet the PHAs criteria.
In outer space, the orbits of { the} Anza group form a structure { resembling} a typical meteoroid stream, as illustrated in Fig.~\ref{fig:Anza23}. The names of the members of this association are listed in Table~ \ref{tab:Anza}.

{ The asteroid} Anza was included by \citet{1991Icar...89...14D} as a member of association II in his Table II. However, the other members of this group ((3551) Verenie, (496817) 1989VB, (3908) Nyx and (2202) Pele)  were not identified as the members of our Anza association. Therefore our finding { does} not confirm Drummond's result. 
\begin{table}
\scriptsize
\begin{center}
\caption{Codes of the members of Anza (2061) association. $89$ members were identified by { the} $D_H$ function, { the} single linkage cluster analysis algorithm and the orbital similarity threshold $D_t=0.047386$.
$38$ objects in bold are of { the} Chelyabinsk class.}
\begin{tabular}{llllll}
\hline
    2061        & \sz{100085    } & \sz{162183    } &     222008      & \sz{1993UA    } & \sz{2001TY1   } \\ 
\sz{2002VU17  } & \sz{2004SS    } & \sz{2005PA17  } &     2005UC3     & \sz{2005VT2   } &     2006SN131   \\ 
\sz{2006UY16  } & \sz{2006VT13  } &     2007PP9     & \sz{2007RG2   } & \sz{2007SS1   } &     2007TD1     \\ 
    2007UO6     &     2007VD8     &     2008RE1     &     2008SF8     & \sz{2008UW99  } &     2008VB4     \\ 
\sz{2009QH34  } & \sz{2009SK    } &     2009SV19    &     2009SX17    &     2009UL20    & \sz{2010SK15  } \\ 
    2010SU15    & \sz{2011RJ1   } & \sz{2011SQ32  } &     2011YQ10    &     2012PZ19    &     2012RH15    \\ 
    2012RW16    &     2012TE53    & \sz{2012UC69  } &     2012VF77    &     2013PH3     & \sz{2013RO21  } \\ 
\sz{2013RT43  } & \sz{2013SQ19  } &     2013TD      & \sz{2013UQ5   } & \sz{2013VH13  } &     2013VP13    \\ 
    2014SE145   &     2014ST1     &     2014SU143   &     2014TH17    & \sz{2014TZ17  } &     2014UY57    \\ 
\sz{2014WG70  } &     2015MW130   & \sz{2015PT227 } &     2015TP21    &     2015US51    &     2015VC65    \\ 
    2015VK64    &     2016NA1     &     2016PR26    &     2016QQ44    &     2016RD20    &     2016TC57    \\ 
\sz{2016TL2   } &     2016TQ18    & \sz{2016TS    } & \sz{2016TT55  } &     2016TT92    & \sz{2016UD26  } \\ 
\sz{2016UH101 } &     2017QW32    &     2017SG16    & \sz{2017SQ12  } & \sz{2017TE1   } &     2017TL2     \\ 
\sz{2017TM2   } &     2017UM2     &     2017UU      & \sz{2017VA    } & \sz{2018RC4   } &     2018RR4     \\ 
    2018SC1     & \sz{2018TB    } &     2018TY4     &     2018VO5     &     2018VP6     &  \\ 
\hline
\label{tab:Anza}
\end{tabular}
\end{center}
\normalsize
\end{table}
\subsection{(12923) Zephyr association}
\begin{figure}
\begin{center}
\vbox{
\includegraphics[width=0.30\textwidth]{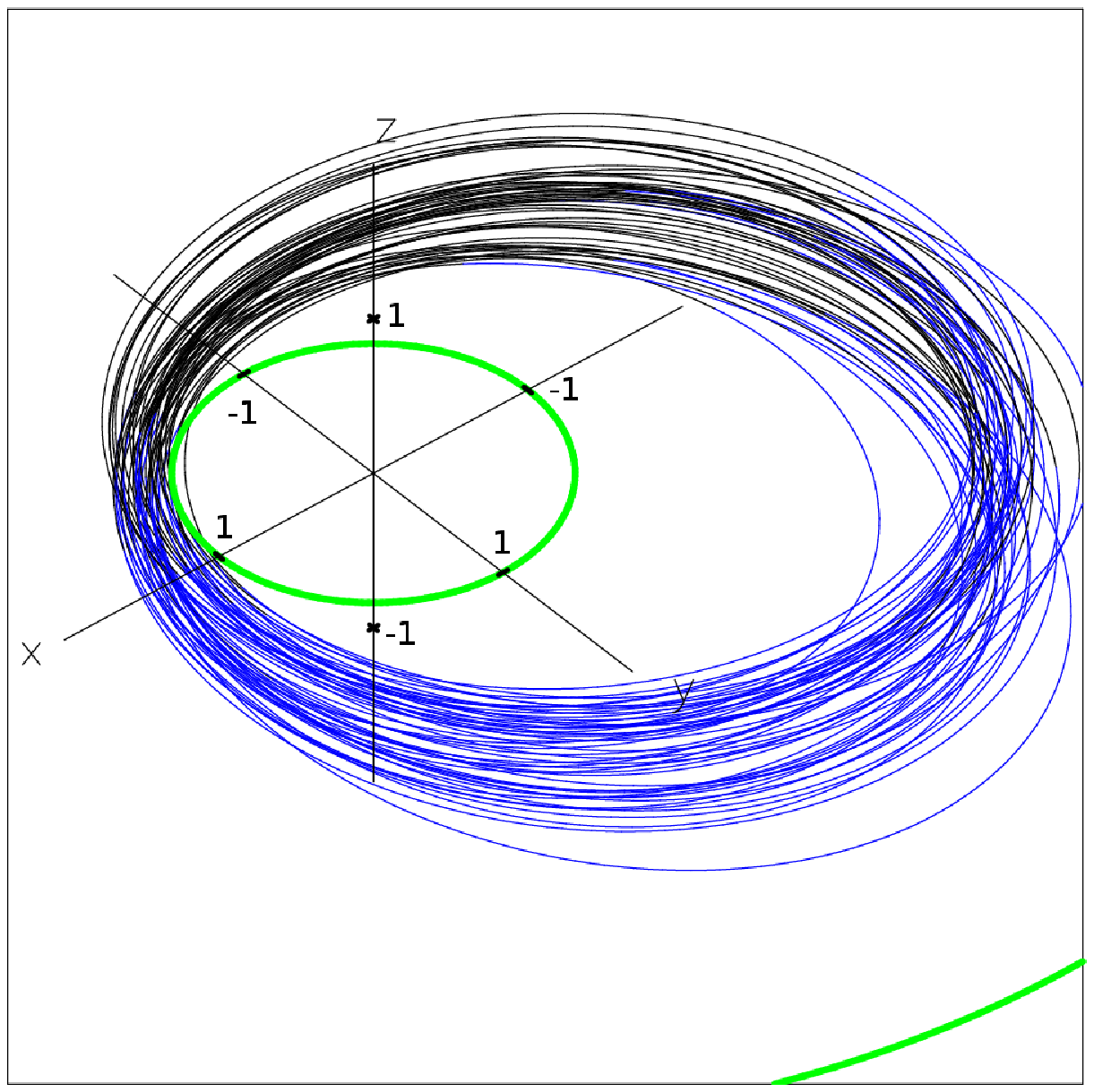}   
\\
\includegraphics[width=0.30\textwidth]{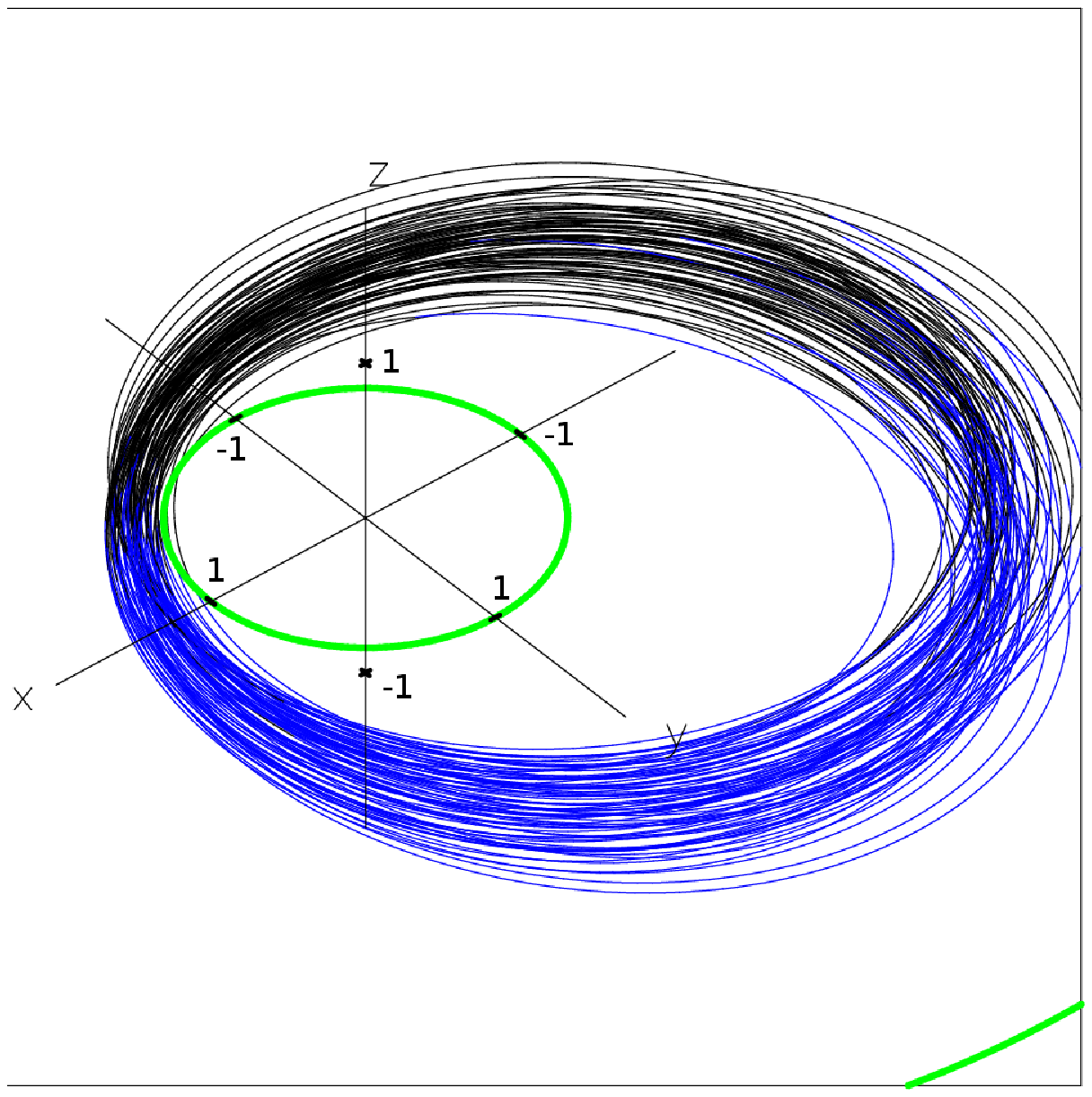}
}
\end{center}
\caption{Two plots of the Zephyr (12923) association { orbits}. { Top} --- $51$ orbits observed before AD $2012$, { bottom} --- $81$ orbits observed { between} $2012$--$2019$. The drawings are similar to each other. { Earth's} orbit is plotted { in green}.}
\label{fig:Zephyr178}
\end{figure}

Zephyr group (DH/178, H=$15.7^m$, $d=2060$ [m]) was identified by { the} $D_H$ function and $D_t=0.047304$. It consists of $51$ orbits observed up to AD $2012$ and $81$ observed since $2012$. In our previous search \citep{2015HiA....16..474J} this association { had been} named NEA  Masaakikoyama (13553). { The} $132$ orbits (see Tab.~\ref{tab:Zephyr}) { form a structure resembling a meteoroid stream} (see Fig.~\ref{fig:Zephyr178}), however not as coherent one as { in { the} case of} Anza association. 
The smallest MOID for this group is $0.000157$ [au] or $0.0611$ [LD]. We found it for { the asteroid} 2018SM ($H=29.4^m$). { The} parameters of { the} theoretical radiant of this NEA are:  $\alpha_G=217^\circ$, $\delta_G=20^\circ$,  $V_G=10.2$ [km/s]. { The} potential meteor-bolide activity of this radiant falls on September 17th. The maximum MOID in Zephyr group equals $0.2768$ [au] { with} the average one { of} $0.1417$ [au]. $14$ members are of { Chelyabinsk} class, $6$ objects meet the PHA criteria.
\begin{table}
\scriptsize
\begin{center}
\caption{Zephyr (12923) association. $132$ members were identified by { the} $D_H$ function, { the} single linkage cluster analysis algorithm and the orbital similarity threshold $D_t=0.047304$. $14$ objects in bold are of { the}  Chelyabinsk class.}
\begin{tabular}{l l l l ll}
\hline
\sz{12923     } &     13553       &     27031       &     159856      &     161999      & \sz{212546    } \\ 
    235086      &     373393      &     416584      &     452314      &     468741      &     481984      \\ 
\sz{2000PP9   } &     2000RK60    &     2001NJ6     &     2001QD34    &     2001SO263   &     2002RB      \\ 
\sz{2003MU    } &     2003QA30    &     2003SA85    &     2003SV159   &     2004PD97    &     2004PE20    \\ 
\sz{2004QJ13  } &     2004RS25    & \sz{2004SE26  } &     2005MX1     &     2005OK3     &     2005TL      \\ 
    2006HB      &     2006JU      &     2006KQ21    &     2007RS12    &     2007RZ19    &     2007TS68    \\ 
    2009QC      &     2009QR8     &     2010JV153   &     2010KK127   &     2010RP64    &     2010RS3     \\ 
\sz{2011HD63  } &     2011JP29    &     2011OC18    &     2011OE16    &     2011OV17    &     2011PC2     \\ 
\sz{2012HG31  } &     2012NP      &     2012OU5     &     2012PP24    &     2012QX17    &     2013KF6     \\ 
    2013NL10    &     2013PC39    &     2013PD26    &     2013PU13    &     2013PV6     &     2013QB11    \\ 
    2013QK6     &     2013RO30    &     2014JR55    & \sz{2014MJ26  } & \sz{2014MP5   } &     2014ON339   \\ 
\sz{2014OQ207 } &     2014OR338   &     2014PT59    &     2014QE391   & \sz{2014QE434 } &     2014QK390   \\ 
    2014QM362   &     2014SV141   &     2015KK122   &     2015MJ130   &     2015MU59    &     2015NV13    \\ 
    2015OH22    &     2015OJ22    &     2015PP227   &     2015PV227   &     2015PW56    &     2015QS8     \\ 
    2015QU8     &     2015RB36    &     2015RC2     &     2015RC37    &     2015RF83    &     2015RP2     \\ 
    2015TS237   &     2016GN220   &     2016LS9     &     2016MF1     &     2016NC39    &     2016NM39    \\ 
    2016NQ16    &     2016PB2     &     2016PH66    &     2016PW38    &     2016PW78    &     2016QR44    \\ 
    2016RC41    & \sz{2016RK17  } &     2016UD101   &     2017MC5     &     2017MY2     & \sz{2017OF7   } \\ 
    2017OG7     &     2017OU67    &     2017PC27    &     2017QA17    &     2017QK33    &     2017QO16    \\ 
    2017QO32    &     2017QU17    &     2017RC15    &     2017SQ21    &     2018JC3     &     2018LS15    \\ 
    2018MR6     &     2018PJ21    &     2018PL22    &     2018PN23    &     2018PO9     &     2018PO10    \\ 
    2018PQ24    &     2018PW21    &     2018PY21    &     2018RK3     &     2018RV1     &     2018SM      \\ 
\hline
\label{tab:Zephyr}
\end{tabular}
\end{center}
\normalsize
\end{table}
\subsection{(4179) Toutatis association}
\begin{figure}
\begin{center}
\vbox{
\includegraphics[width=0.30\textwidth]{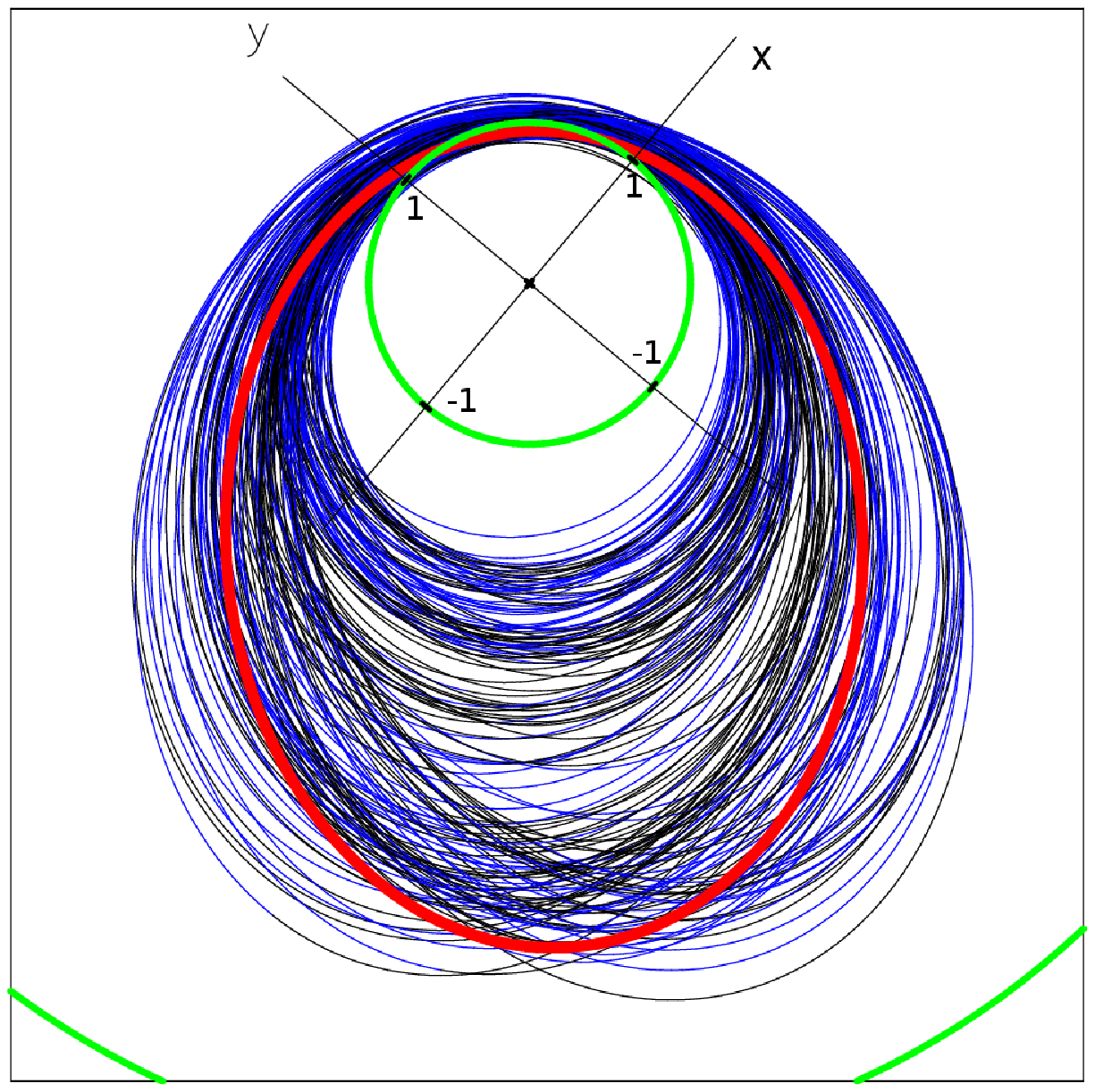}    
\\
\includegraphics[width=0.30\textwidth]{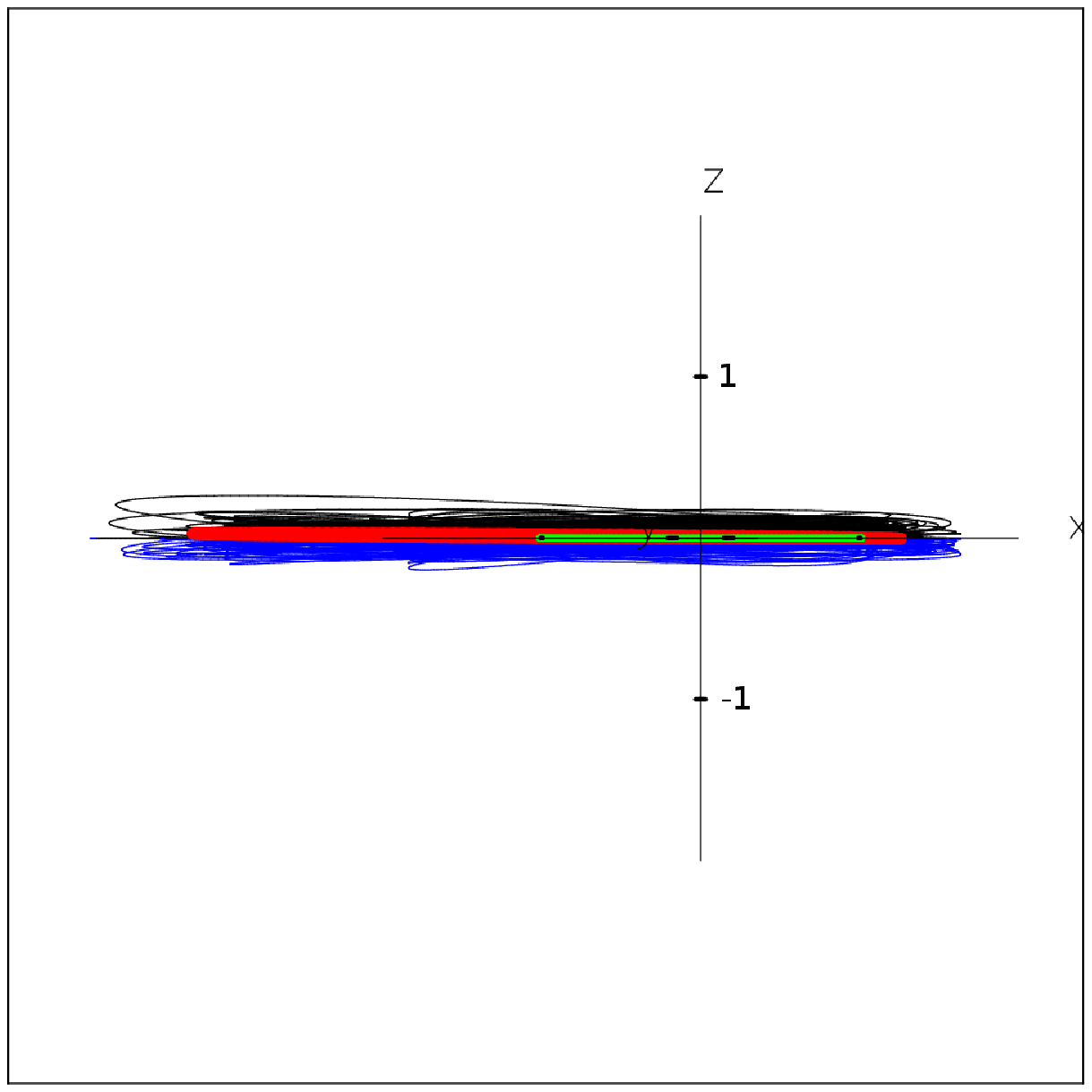}
}
\end{center}
\caption{Two projections of Toutatis { association orbits} (4179). { Top} --- the ecliptic projection, { bottom} --- the projection { perpendicular} to the ecliptic. The group was identified by { the} $\rho_1$ function, the orbit count { is} $173$. { The} Toutatis orbit is plotted { in red}. { Earth's} orbit is plotted { in green}.}
\label{fig:Toutatis}
\end{figure}

{ Toutatis} association (internal code $\rho_1$/$63$) { was} named after the most massive object in the group, $H$$=$$15.2^m$, size  $4750$x$1950$ [m]. 
It was identified using { the} $\rho_1$ and $D_H$ functions, { providing} $173$ (see Tab.~\ref{tab:Toutatis}) and $211$ members, respectively. The group { resembles a} dispersed but still coherent meteoroid stream. The { orbits'} inclinations { are} small (up to $6.7^\circ$); many orbits closely { approach} { Earth's} orbit at ascending and descending nodes. Toutatis orbital inclination equals $0.45^\circ$, its orbit occupies central position in the group, see Figure \ref{fig:Toutatis}. The minimum, maximum and mean MOID for this group are: $0.000102$ [au] or $0.039695$ [LD], $0.182691$ [au], $0.036038$ [au], respectively. The minimum is for 2017SR2 NEA and the radiant parameters are  $\alpha_G=323^\circ$, $\delta_G=-18^\circ$, geocentric velocity $V_G=9.5$ [km/s]. Potential meteor-bolide activity of this radiant falls on September 22nd. $71$ members of the Toutatis association are of { Chelyabinsk} class.
%
\begin{table}
\scriptsize
\begin{center}
\caption{Toutatis (4179) association. $173$ members identified by { the} $\rho_1$ function, { the} single linkage cluster analysis algorithm and the orbital similarity threshold $D_t=0.053066$. { In}  bold, $71$ { association members which} are of { Chelyabinsk} class.}
\begin{tabular}{l l l l ll}
\hline
\sz{4179      } & \sz{65803     } &     429382      & \sz{440212    } &     455322      & \sz{1997UR    } \\ 
\sz{2000SB45  } & \sz{2000TE2   } &     2001UD18    & \sz{2002TP69  } & \sz{2003TO9   } & \sz{2003WY153 } \\ 
\sz{2004XG29  } & \sz{2005AZ28  } & \sz{2005SK26  } &     2005TD49    &     2005UF7     & \sz{2005UN    } \\ 
    2005VA      & \sz{2006UA216 } &     2006UE185   & \sz{2006UN    } &     2006VU2     & \sz{2006YP44  } \\ 
\sz{2007TA23  } &     2007TT24    & \sz{2007TU68  } & \sz{2007VD138 } &     2007VH184   & \sz{2007WC    } \\ 
\sz{2007YY59  } &     2008SV7     &     2008WF14    &     2008WF96    & \sz{2009QZ34  } & \sz{2009RG2   } \\ 
\sz{2009RV1   } &     2009TF8     &     2009UK1     &     2009UN1     &     2009VT1     & \sz{2009WG106 } \\ 
    2009WX7     &     2010UA7     &     2010UO7     &     2010VA40    & \sz{2010VA140 } &     2010VC140   \\ 
    2010VF139   &     2010VQ139   &     2010VR139   &     2010VV21    &     2010VW194   &     2010VY71    \\ 
\sz{2010WD9   } & \sz{2010XW58  } &     2011CR4     & \sz{2011QS49  } & \sz{2011SL189 } &     2011UB      \\ 
\sz{2011YQ1   } & \sz{2012SJ32  } &     2012TC4     &     2012TQ20    & \sz{2012VB37  } &     2012VF26    \\ 
    2012VL5     & \sz{2012XB133 } &     2012XH      & \sz{2013RC32  } &     2013RS43    & \sz{2013TO69  } \\ 
\sz{2013US8   } &     2013UU1     &     2013VR12    & \sz{2013XY9   } & \sz{2014OM339 } &     2014SX141   \\ 
    2014TR      & \sz{2014UA57  } &     2014UH57    &     2014UK114   &     2014WF364   & \sz{2014WG200 } \\ 
    2014WN69    &     2014WX4     & \sz{2015HV182 } & \sz{2015RN35  } &     2015SJ17    &     2015TC145   \\ 
    2015TC179   &     2015TR238   &     2015TZ143   & \sz{2015UC    } & \sz{2015UH    } &     2015VA1     \\ 
\sz{2015VG150 } &     2015VH1     & \sz{2015WF    } & \sz{2015XE    } &     2015XG1     & \sz{2015XM169 } \\ 
\sz{2015XR378 } & \sz{2016PR38  } &     2016SO2     &     2016TL94    &     2016TW19    &     2016UD42    \\ 
    2016UG57    &     2016UY5     &     2016VL2     &     2016VQ4     &     2016WE      &     2016WM7     \\ 
    2016WQ1     &     2016XL23    & \sz{2016XQ1   } & \sz{2016YO    } & \sz{2017QU1   } &     2017RO16    \\ 
\sz{2017RU2   } & \sz{2017RU17  } & \sz{2017RW2   } &     2017SJ14    & \sz{2017SK14  } &     2017SL2     \\ 
    2017SR2     &     2017ST10    &     2017TK4     &     2017TQ5     &     2017UG3     &     2017UJ44    \\ 
    2017UO1     &     2017UQ5     &     2017US7     &     2017UT5     &     2017VC13    & \sz{2017VD15  } \\ 
    2017VE13    &     2017VG13    &     2017VG34    & \sz{2017VM2   } &     2017VO14    &     2017VX14    \\ 
    2017WL2     &     2017WL28    &     2017WP16    & \sz{2017WW13  } & \sz{2017XF2   } & \sz{2017YK5   } \\ 
\sz{2018AY    } & \sz{2018BP    } &     2018RK6     & \sz{2018RM4   } &     2018RQ4     &     2018SB1     \\ 
    2018TM3     &     2018TN3     & \sz{2018TS2   } &     2018TV      &     2018VB5     &     2018VG6     \\ 
    2018VK3     &     2018VN1     &     2018VR7     &     2018VX4     & \sz{2018VX5   } &     2018WQ      \\ 
\sz{2018WV    } &     2018WZ2     &     2018XF2     & \sz{2018XX4   } & \sz{2019AS2   } &  \\ 
\hline
\label{tab:Toutatis}
\end{tabular}
\end{center}
\normalsize
\end{table}

{ The} hypothesis that Toutatis is a member of the NEA group, in the context of Taurids Complex, was raised by \citet{1993MNRAS.264...93A,1993DPS....25.3206S, 1994ASPC...63...97S}. In \citet{ 1994ASPC...63...97S} choosing Southworth and Hawkins D-function \citep{1963SCoA....7..261S} and $D_t=0.25$, Toutatis was listed { with} $22$ other NEAs as having orbits { similar} to { those of} the Taurid meteoroid complex. In our search however, none of the $22$ orbits was found among our Toutatis group. 

{ Speculating} if Toutatis is a member of the NEA family, Steel wrote {\it 'Asteroid (4179) Toutatis is of some topical interest since it had a near miss of the Earth in 1992 December, radar images showing it to have a bifurcated structure. This might be understood in terms of a hierarchical disintegration, several sub-units having already separated, leaving two still in contact. This interpretation would be bolstered by the identification of other NEAs on similar orbits as Toutatis, indicating that a family has been produced
by the break-up of a larger object ... '}. On December 2012, { a} close observation of Toutatis { performed} by Chang'e-2 spacecraft, \citep{2013NatSR...3E3411H} { confirmed} Steel's speculations. The bifurcated configuration implies { that Toutatis is} a contact binary. { Other} details { from the} images imply Toutatis may { have} a rubble-pile structure \citep{2013NatSR...3E3411H}. 

{ Therefore} our finding is in favour of { a} catastrophic origin of { the asteroid} Toutatis, and perhaps the { entire} association. However, it is not possible to { fully} answer questions { that are} related with the origin of { the asteroid} Toutatis association after { using solely} a cluster analysis.
\subsection{(25143) Itokawa  association}
This group (DH/227) was identified by { the} $D_H$, $\rho_1$ as well as by { the} $\rho_2$ function. The numbers of { the} identified NEAs were $94, 128, 106$, respectively. One of the most massive object in the group is { the asteroid} Itokawa, (25143), $H=18.95^m$, size $520$x$270$x$230$ [m]. $94$ members of the group were identified with { the} $D_H$ and $D_t=0.047529$.

\begin{figure}
\begin{center}
\vbox{
\includegraphics[width=0.30\textwidth]{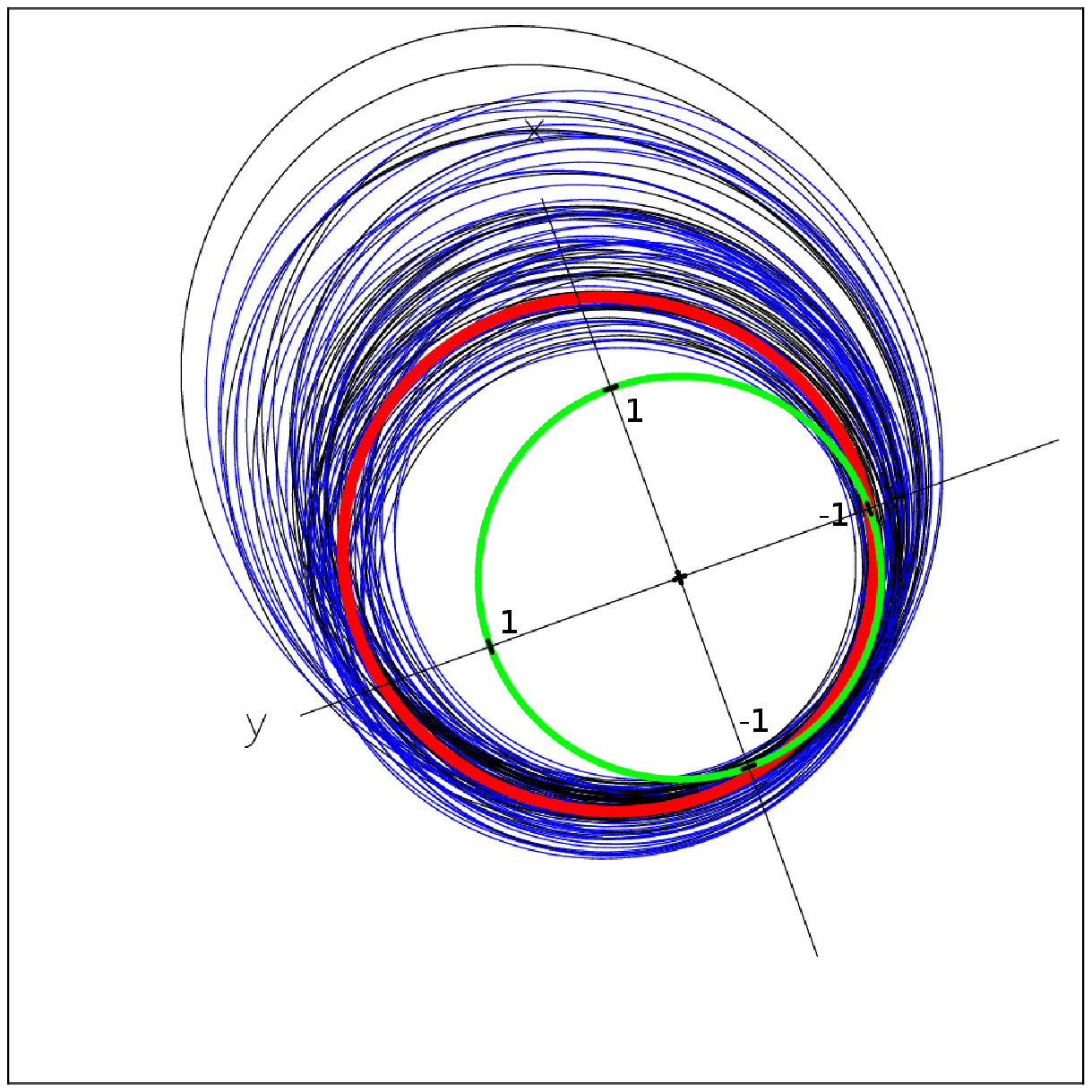}   
\\
\includegraphics[width=0.30\textwidth]{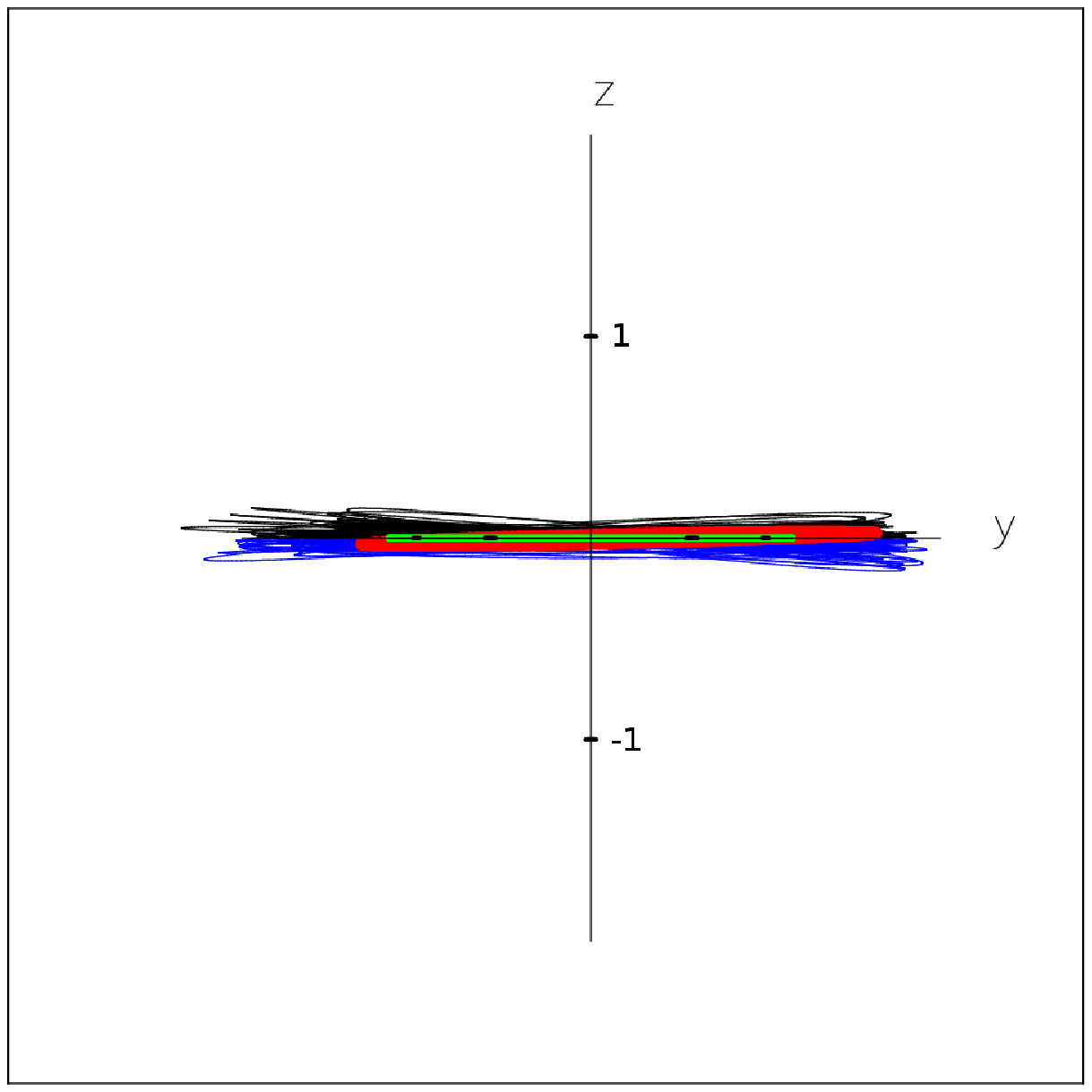}
}
\end{center}
\caption{Two projections of { the} Itokawa  (25143) association { orbits}. { Top} --- 3-D projection from above the ecliptic, { bottom} --- the  projection { perpendicular} to the ecliptic. The group was identified by { the} $D_H$ function, it { consists of}  $94$ orbits. Itokawa orbit is plotted in red, { Earth's} orbit is plotted in green.}
\label{fig:Itokawa}
\end{figure}

The group resembles { a coherent} meteoroid stream. The inclinations of the orbits are small (up to $6.5^\circ$), many orbits approach { Earth's} orbit {closely} at both nodes. Itokawa orbital inclination equals $1.62^\circ$, { its} orbit approaches { Earth's} orbit within $0.013$ and $0.023$ [au] on 28th June and 4th April, respectively. It occupies { an} internal position in the group, see Fig.~\ref{fig:Itokawa}. The minimum, maximum and mean MOIDs of this group are: $0.000037$ [au], $0.085742$ [au], $0.022713$ [au], respectively. The smallest MOID { of} $0.014399$ [LD] was found for 2017FU102, $H$$=$$28.67^m$, $d$$=$$6$ [m]. { This} MOID is smaller than { Earth's} radius but, fortunately, at the date when our planet was close to this point the asteroid was far away on its orbit. This small Earth MOID { was} calculated using the orbital elements corresponding to the osculating epoch $2455400.5$ [TDB]. The orbit of 2017FU102 is not well known, the data-arc span is { only} $11$ days. 

{ The} theoretical radiant parameters related to { the asteroid}  2017FU102 are $\alpha_G$$=$$178^\circ$, $\delta_G$$=$$16^\circ$ and $V_G$$=$$7.29$ [km/s] the date of possible activity of this radiant is April 3rd.
{ Forty}  members of this association satisfy { the} conditions of { the} { Chelyabinsk} class, see Table \ref{tab:Itokawa}.
%
\begin{table}
\scriptsize
\begin{center}
\caption{Itokawa (25143) association. $94$ members identified by { the} $D_H$ function, a single linkage cluster analysis algorithm and the orbital similarity threshold $D_t=0.047529$. $40$ members of { the} Itokawa association, including { the asteroid} Itokawa, are of { the} { Chelyabinsk} class.}
\begin{tabular}{l l l l ll}
\hline
\sz{25143     } & \sz{89136     } &     141018      &     2003DW10    &     2004DK1     & \sz{2004HD    } \\ 
\sz{2004OW10  } & \sz{2006DN    } & \sz{2006EY    } &     2006HX30    &     2006JT41    & \sz{2006KL103 } \\ 
    2007EG88    &     2007HB15    & \sz{2007LT    } & \sz{2008CC71  } & \sz{2008HB2   } &     2008LE      \\ 
    2009DN4     &     2009FH      &     2009FK      &     2009HH21    & \sz{2010CE55  } &     2010HF      \\ 
\sz{2010JW39  } & \sz{2011EC    } &     2011FT9     &     2011FU9     & \sz{2011GR59  } & \sz{2012EB2   } \\ 
\sz{2012EM8   } & \sz{2012FV23  } &     2012GP1     & \sz{2012HB2   } & \sz{2012JO4   } &     2012KT12    \\ 
    2013ED68    &     2013EN20    &     2013EN89    &     2013EN126   & \sz{2013FB8   } & \sz{2013JF1   } \\ 
    2013JT28    & \sz{2014DG80  } & \sz{2014EO12  } &     2014FZ      &     2014GV48    &     2014HA199   \\ 
    2014HR198   &     2014HS4     &     2014HT46    &     2014HV197   &     2014HY197   &     2014KQ84    \\ 
\sz{2015CW13  } & \sz{2015DK200 } &     2015FC345   &     2015FK120   &     2015HB117   &     2015KD57    \\ 
    2015KK120   & \sz{2015MB54  } & \sz{2016CY135 } & \sz{2016EF86  } &     2016EQ1     &     2016FN13    \\ 
    2016GH1     &     2016GH3     & \sz{2016HO    } &     2016JH      &     2016JM      & \sz{2016JN6   } \\ 
    2016NL39    & \sz{2016NV22  } &     2017FA159   &     2017FU102   &     2017FW128   &     2018CZ13    \\ 
\sz{2018DN1   } & \sz{2018DY3   } &     2018ED9     & \sz{2018EM4   } &     2018HO      &     2018PD9     \\ 
\sz{2018PR23  } & \sz{2019CL2   } & \sz{2019ES    } & \sz{2019ES2   } & \sz{2019FH1   } &     2019GC      \\ 
    2019GP5     & \sz{2019GT1   } &     2019JA1     &     2019JP1     &                 &   \\ 
\hline
\label{tab:Itokawa}
\end{tabular}
\end{center}
\normalsize
\end{table}

In November 2005, { the spacecraft} Hayabusa landed on { the asteroid} 25143 Itokawa { and}  captured { dust} particles which were { then} analysed in detail. In \citet{2011Sci...333.1113N} the authors { noted that} {\it `Mineral chemistry indicates that the majority of { the surface} regolith particles suffered long-term thermal annealing and subsequent impact shock, suggesting that Itokawa is an asteroid made of reassembled pieces of the interior portions of a once larger asteroid`}. 
Following this ascertainment, \citet{2011PASJ...63L..73O} { conducted} a search for meteoroids originating from this NEA. { They} found that the fireball MORP 172\footnote{The fireball photographed by the Canadian   Meteorite  Observation and  Recovery Project.} { reveals a  dynamic} similarity to { the asteroid} Itokawa. Using the \"{O}pik variables $U, cos(\theta)$, \citep[see e.g.][]{1999MNRAS.304..743V} they claim { that there is} {\it ''a strong dynamical relation (thus genetic) between Itokawa and MORP 172``}. 
{ Therefore,} our finding { concerning} { the} Itokawa association is consistent with {   the ascertainments of} \citet{2011Sci...333.1113N} and \citet{2011PASJ...63L..73O}.
\subsection{(65803) Didymos  association}
\begin{figure}
\begin{center}
\vbox{
\includegraphics[width=0.30\textwidth]{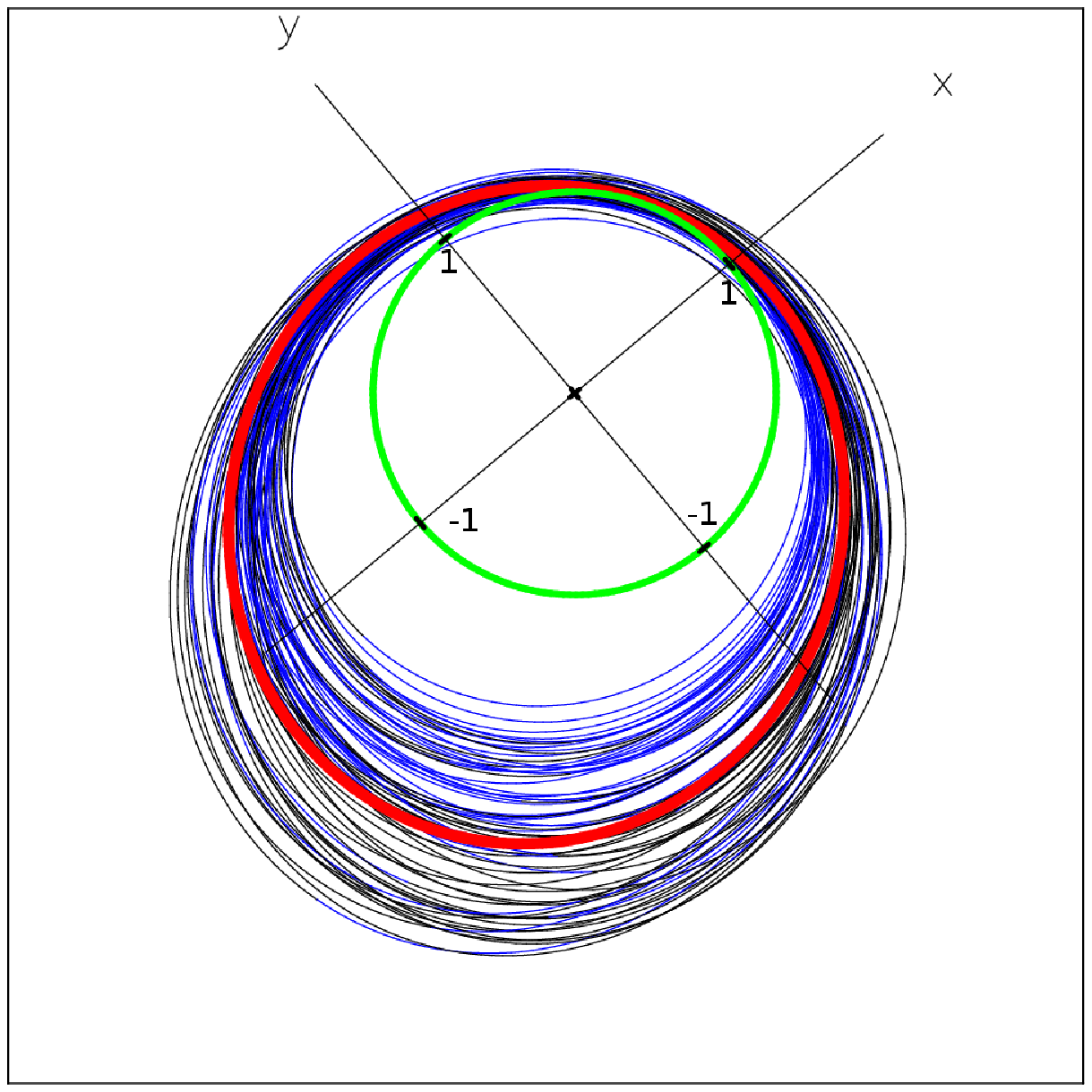} 
\\
\includegraphics[width=0.30\textwidth]{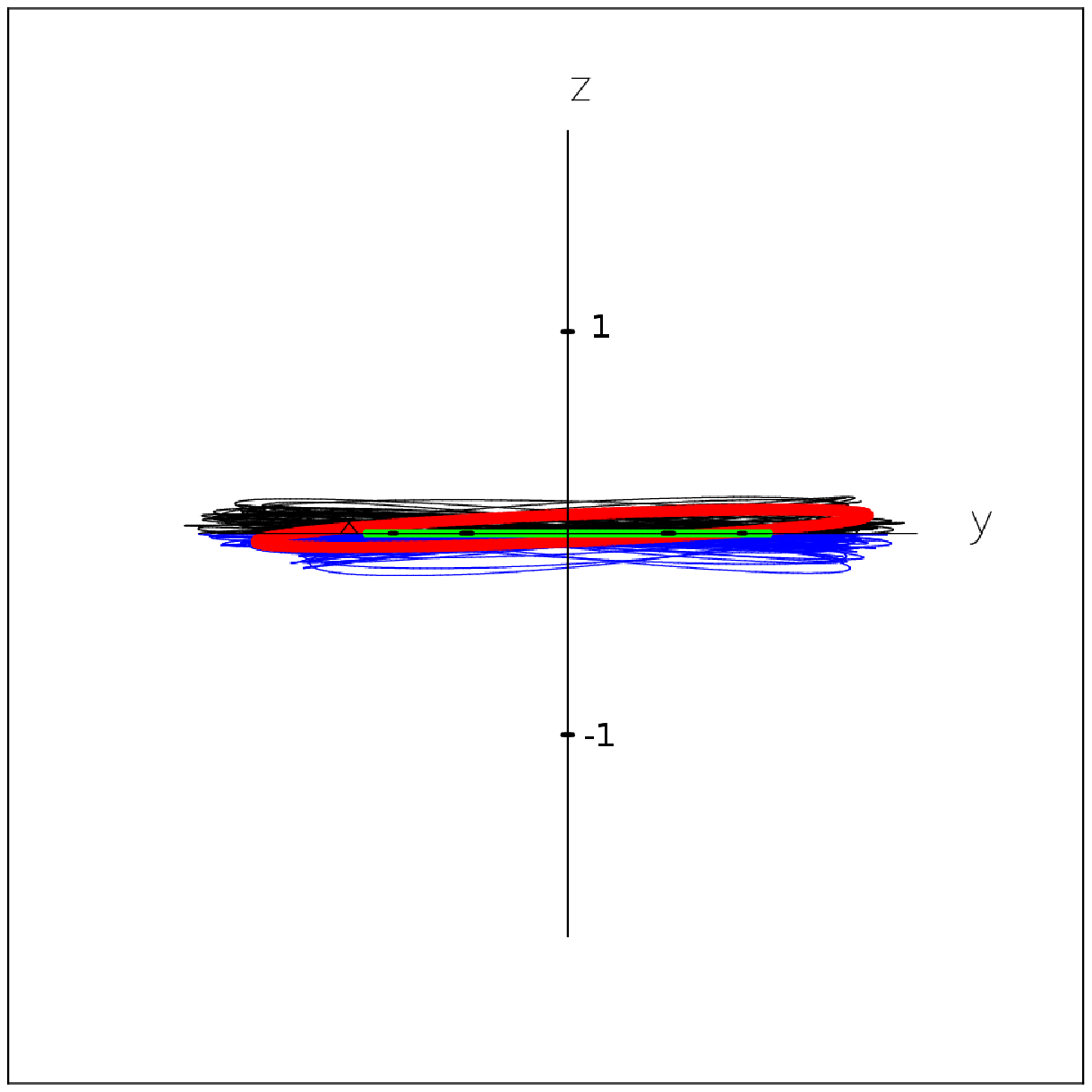}
}
\end{center}
\caption{Two projections of { the} (65803) { Didymos} association { orbits}. { Top} --- the ecliptic projection, { bottom} --- the projection { perpendicular} to the ecliptic. The group was identified by { the} $\rho_2$ function, { there are} $70$ orbits. Didymos orbit is in red, { Earth's} orbit is in green. }
\label{fig:Didymos}
\end{figure}

{ The} Didymos group ($\rho_2/309$) was { identified} by { the} $D_H$, and $\rho_2$ functions { with} $20$ { and} $70$ members respectively. One of the most massive { objects} in the group is { the asteroid} Didymos (65803), $H$$=$$17.98^m$, size $800$ [m]. { Seventy} members of the group were identified by { the} $\rho_2$ function and $D_t$$=$$0.056566$. { The} Didymos association asteroids { are listed} in { Table}~\ref{tab:Didymos}, { and} their orbits are illustrated in Fig. \ref{fig:Didymos}.

The minimum, maximum and mean Earth MOIDs of this group are: $0.000293$ [au], $0.11116$ [au], $0.02112$ [au], respectively. The smallest { MOID calculated} equals $0.114027$ [LD] { and was} found for 2016WQ1, $H$$=$$27.89^m$. { The} theoretical geocentric radiant parameters of this object are $\alpha_G$$=$$284^\circ$, $\delta_G$$=$$-22^\circ$ and $V_G$$=$$7.63$ [km/s] { and} the date of possible activity of the radiant is November 18th. { Thirty-one} members of this group { meet} { the} Chelyabinsk class conditions.
\begin{table}
\scriptsize
\begin{center}
\caption{{ (65803) Didymos} association. $70$ members were identified by { the} $\rho_2$ function, { the} single linkage cluster analysis algorithm and the orbital similarity threshold $D_t=0.056566$. $31$ members of this associations are of { the} Chelyabinsk class.}
\begin{tabular}{l l l l ll}
\hline
\sz{65803     } & \sz{440212    } & \sz{2000TE2   } &     2001UD18    & \sz{2005SK26  } & \sz{2005UN    } \\ 
\sz{2005WK56  } & \sz{2006UA216 } & \sz{2006UN    } & \sz{2007TA23  } &     2007TT24    &     2008SV7     \\ 
    2008WF14    & \sz{2009QZ34  } & \sz{2009RG2   } & \sz{2009WG106 } &     2009WX7     &     2010RD      \\ 
\sz{2010RF31  } &     2010UO7     & \sz{2010VA140 } &     2010VQ139   &     2010VR139   &     2010VW194   \\ 
\sz{2010WD9   } & \sz{2011QS49  } &     2011UB      & \sz{2012VB37  } &     2012VL5     &     2012XH      \\ 
    2013RS43    & \sz{2013US8   } & \sz{2013XY9   } &     2014WN69    &     2014WX4     &     2015SJ17    \\ 
    2015TC179   &     2015VA1     &     2015VH1     & \sz{2015XE    } &     2015XG1     & \sz{2015XM169 } \\ 
\sz{2016PR38  } & \sz{2016RY17  } &     2016SO2     &     2016TL94    &     2016UY5     & \sz{2016VK6   } \\ 
    2016WM7     &     2016WQ1     &     2017QO17    & \sz{2017QU1   } &     2017TK4     &     2017TQ5     \\ 
    2017UQ5     &     2017UT5     & \sz{2017VD15  } &     2017VG13    & \sz{2017XF2   } & \sz{2018BP    } \\ 
    2018RQ4     &     2018SB1     &     2018TM3     &     2018TN3     &     2018VB5     &     2018VG6     \\ 
    2018VN1     & \sz{2018XE    } & \sz{2018XX4   } & \sz{2019AS2   } &                &   \\ 
\hline
\label{tab:Didymos}
\end{tabular}
\end{center}
\normalsize
\end{table}
\subsection{(225312) Cyclids-SEAs association}
We adopted the name Cyclids for the NEAs association from \citet{1963SCoA....7..261S} who first found a group of five orbits very similar to that of { Earth's}  among the Super-Schmidt photographic meteoroid data. This discovery was confirmed using { a} larger photographic sample by \citet{1971SCoA...12....1L} and \citet{1999MNRAS.304..751J} and using TV data by \citet{1997A&A...320..631J}. Southworth and Hawkins { indicated} that the Cyclids are quite probably not a meteoroid stream, e.g. their radiant activity spread was very large --- { from} April to October.

In our basic search (using { the} $D_H$ and $D_t$$=$$0.047215$) { we found} a group of $73$ NEAs orbits { which are} very similar to { those of the} Cyclids' { meteoroids} { and} our association is consistent with { the} group mentioned in \citet{1993Natur.363..701C} and \citet{1993Natur.363..704R}. The authors reported an overabundance of NEAs moving on orbits similar to { Earth's} trajectory and called them small-Earth approachers (SEAs). \citet{1993Natur.363..704R} stated that {\it `most of these SEAs have a diameter $d$$<$$50$ [m] and are on low-eccentricity,  low-inclination orbits and semi-major axis not too deviant from unity'}. The sample of SEAs led \citet{1993Natur.363..704R} to suggest { that} there is a near-Earth asteroids belt { with the majority of} { the} members { being} not yet detected. \citet{2008MNRAS.386.2031B} selected $13$ SEAs and studied their dynamical evolution and calculated { the probability} of their impact with { Earth}. It emerged that, except { for} $3$, all remaining SEAs have { much higher} probabilities { of impact} than typical { NEAs}, by up to two orders of magnitude.  

\begin{figure}
\begin{center}
\vbox{
\includegraphics[width=0.32\textwidth]{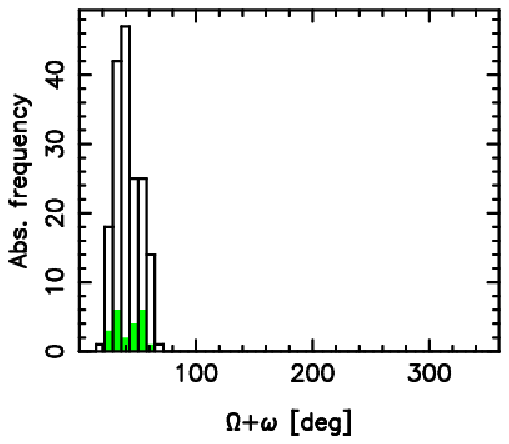}  
\includegraphics[width=0.32\textwidth]{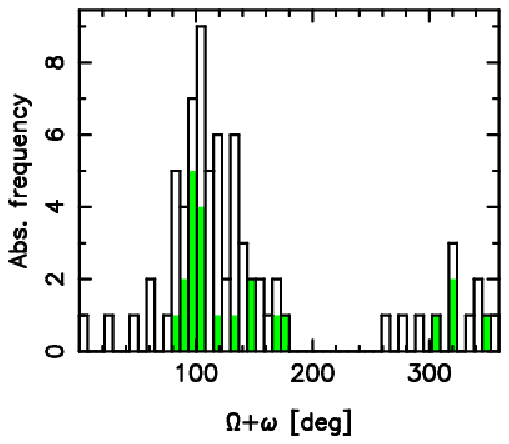}
}
\end{center}
\caption{ { The} histograms of longitude of perihelion ($\varpi=\Omega+\omega$) of { the} Toutatis association (upper panel, $173$  NEAs) and { the} Cyclids-SEAs association (bottom panel, $73$ NEAs). For { the} Cyclids-SEAs { group,} the spread of values is much more considerable and two concentrations of $\varpi$ are well separated, suggesting that this group { consists} of two subgroups. The green { bars} apply to the MOIDs lower { that are} than $0.0025$ [au]. }
\label{fig:longperi}
\end{figure}

Cyclids association { was} found in our previous studies { which are} described in \citet{2011MmSAI..82..310J, 2011epsc.conf...15J}. Using { the} \citet{1963SCoA....7..261S} $D_{SH}$ function, and { the} $D_V$ function defined in \citet{2008EM&P..102...73J}, we found two Cyclids-like groups, \citep[see Table 8 in][]{2011MmSAI..82..310J} which { partly overlapped with}  our Cyclids-SEAs group of $73$ members. Overlooking the distributions of the longitude of perihelion ($\varpi$$=$$\Omega+\omega$) we see that  $\varpi$-distribution of { the} Cyclids-SEAs  differs clearly from { that} of e.g. the Toutatis group, see Fig.~\ref{fig:longperi}. For { the}  Cyclids-SEAs, { the} orientation of the orbit in the plane of motion should be considered { as} less significant.  This { observation} encourages us to be less restrictive  { when} applying the $D_H$ function.  { Thus,} we carried out one more search with a { threshold that was a little} higher, namely with $D_t=0.0502$. As { a} result we identified a group of $129$ Cyclids-SEAs  ($D_H$/955) { and} the names of the members of this association are listed in Table \ref{tab:Cyclids}.
%
\begin{figure}
\begin{center}
\vbox{
\includegraphics[width=0.30\textwidth]{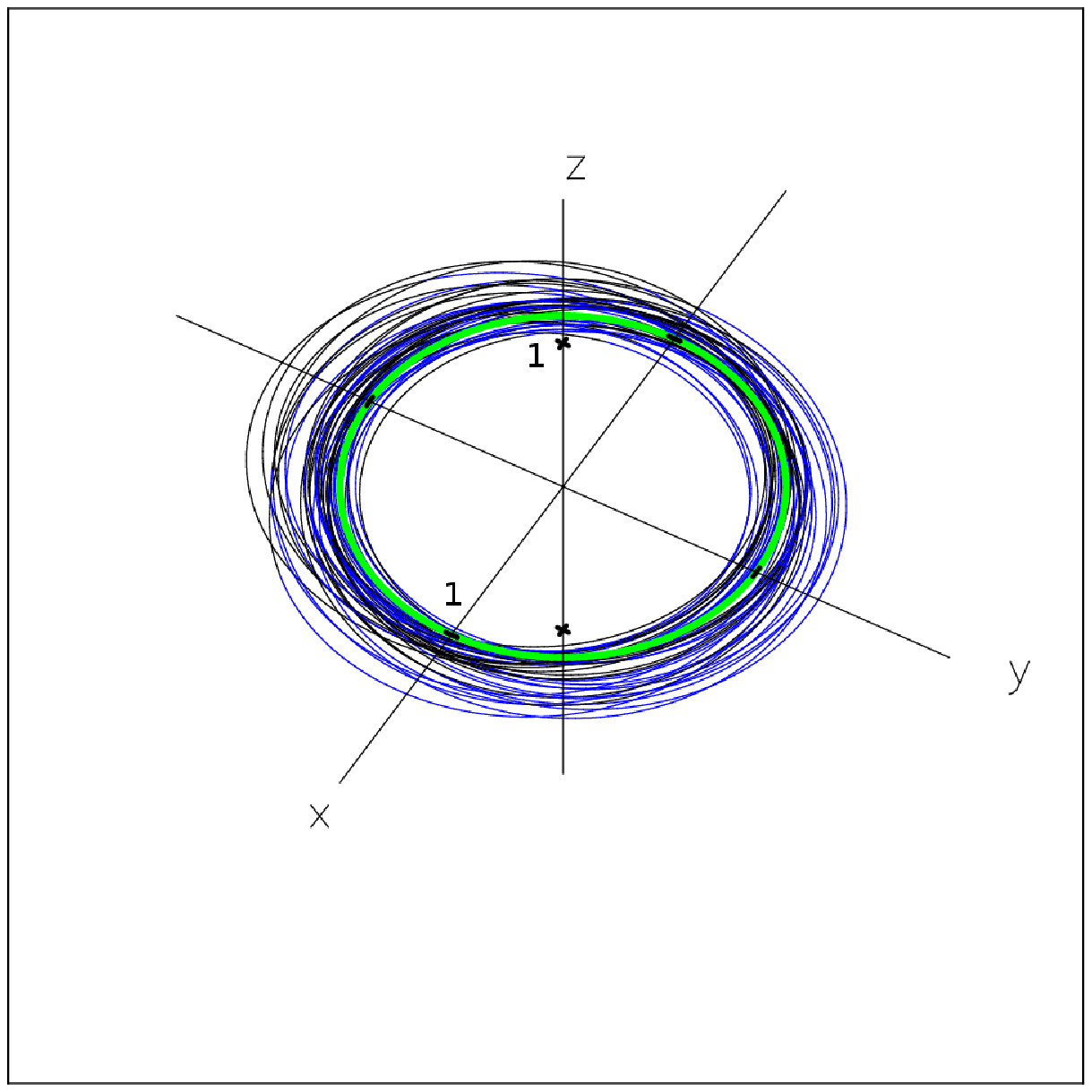}    
\\
\includegraphics[width=0.30\textwidth]{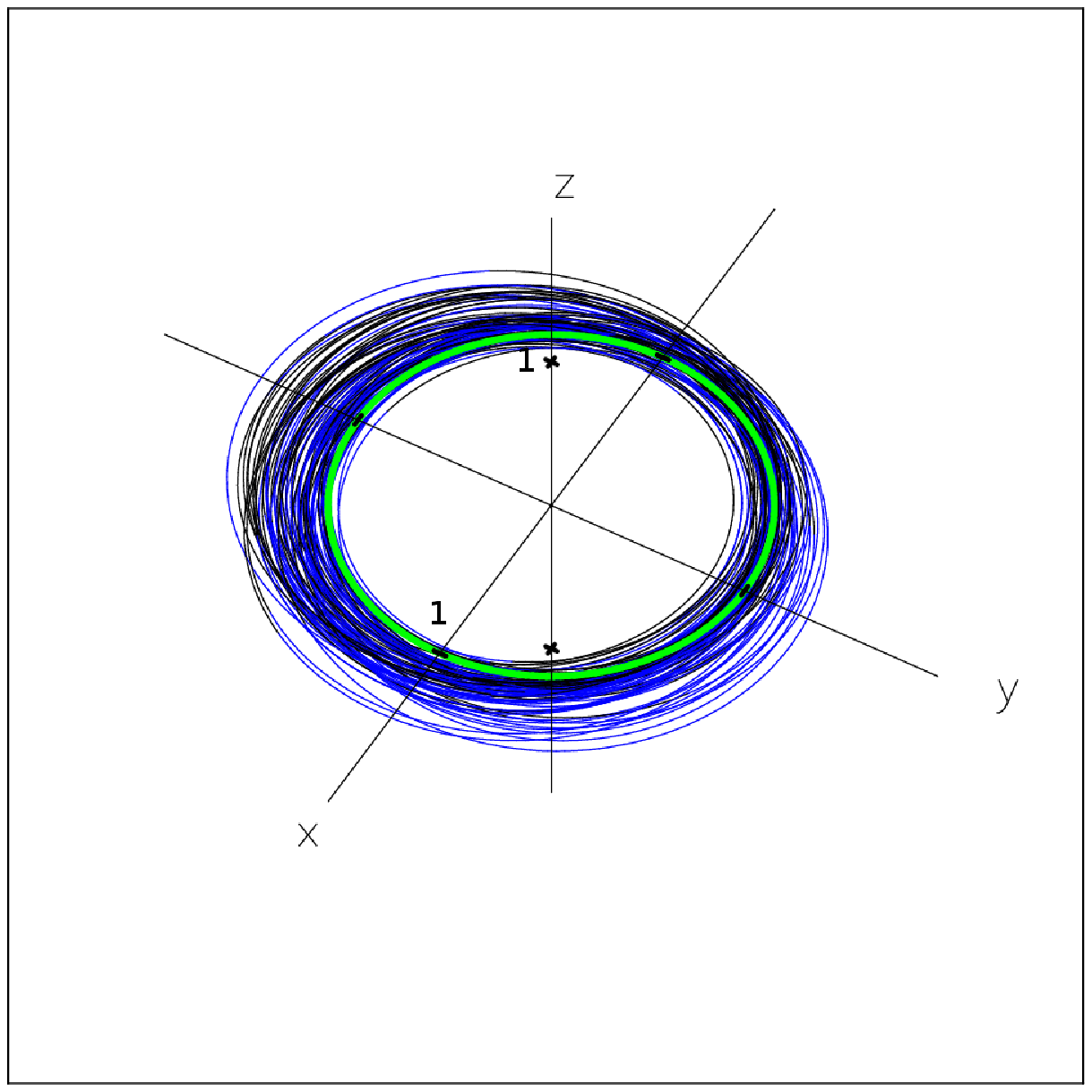}
}
\end{center}
\caption{Cyclids-SEAs association. { Top} --- $45$ orbits observed before AD $2012$, { bottom} ---  $84$ orbits observed between $2012-2019$. { Earth's} orbit is 'immersed' in the association.}
\label{fig:Cyclids}
\end{figure}
\begin{table}
\scriptsize
\begin{center}
\caption{Cyclids-SEAs (225312) association. $129$ members were identified by { the} $D_H$ function, { the} single linkage cluster analysis algorithm and the orbital similarity threshold $D_t$$=$$0.0502$. { The} names of $29$ { Chelyabinsk-class} objects are listed in bold.}
\begin{tabular}{l l l l ll}
\hline
    225312      & \sz{478784    } &     1991VG      & \sz{2000SG344 } &     2001GP2     & \sz{2001QJ142 } \\ 
    2003SM84    &     2004EO20    & \sz{2004WH1   } &     2006BU7     & \sz{2006BZ147 } &     2006JY26    \\ 
    2006RH120   &     2006UQ216   &     2007TF15    &     2007UN12    &     2007VU6     &     2008CM74    \\ 
    2008EA9     &     2008EL68    &     2008HU4     &     2008JL24    &     2008ST      & \sz{2008TX3   } \\ 
    2008UA202   &     2008VC      &     2009BD      & \sz{2009CV    } &     2009DB43    & \sz{2009OS5   } \\ 
    2010AN61    &     2010JW34    &     2010TE55    &     2010UE51    &     2010VQ98    &     2011AA37    \\ 
    2011ED12    &     2011HP24    &     2011UD21    &     2012AQ      & \sz{2012BB14  } &     2012EP10    \\ 
    2012FM35    &     2012FS35    &     2012HG2     &     2012TF79    &     2012UL171   & \sz{2012UW68  } \\ 
    2012XB112   &     2012XM55    & \sz{2013BS45  } &     2013CY      &     2013DA1     &     2013EC20    \\ 
    2013GH66    &     2013LE7     &     2013RZ53    & \sz{2013WA44  } & \sz{2013XY20  } & \sz{2014CH13  } \\ 
    2014HJ197   &     2014KD45    &     2014LJ      &     2014QN266   &     2014UV210   & \sz{2014UY    } \\ 
    2014WA366   &     2014WU200   &     2014WX202   & \sz{2014YD    } &     2015DU      &     2015JD3     \\ 
    2015KE      &     2015KK57    & \sz{2015PL57  } &     2015PS228   &     2015TC25    &     2015VC2     \\ 
    2015VO142   &     2015XZ378   & \sz{2016CF137 } &     2016GK135   & \sz{2016GL222 } &     2016JB18    \\ 
    2016RD34    & \sz{2016TB18  } &     2016TB57    & \sz{2016UE    } &     2016WD7     &     2016WQ3     \\ 
    2016YR      &     2017BG30    & \sz{2017BN93  } &     2017CF32    &     2017FB157   &     2017FF3     \\ 
    2017FJ3     &     2017FL64    &     2017FT102   &     2017FW90    & \sz{2017KP27  } &     2017RL2     \\ 
\sz{2017RP2   } & \sz{2017SV19  } &     2017TA6     &     2017TP4     &     2017UY4     &     2017YK14    \\ 
    2017YS1     &     2018AJ3     &     2018BC      &     2018BJ6     & \sz{2018CQ3   } &     2018CU14    \\ 
    2018DT      & \sz{2018LQ2   } & \sz{2018PM28  } &     2018PU23    &     2018TG6     &     2018UE1     \\ 
    2018VO9     &     2018WV1     &     2019AH13    &     2019AU      & \sz{2019FV2   } &     2019FY1     \\ 
    2019GF1     &     2019GK3     &     2019GL1     &                 &                 &   \\ 
\hline
\label{tab:Cyclids}
\end{tabular}
\end{center}
\normalsize
\end{table}
The association ($D_H$/955) comprises { of}: $13$ Atens, $96$ Apollos and $20$ Amors; among which we found $29$ { Chelyabinsk-class} objects. The mean MOID value equals $0.015817$ [au], the smallest { MOID value} amounts to $0.000058$ [au] ($0.022572$ [LD]) and the greatest { MOID value is} $0.159817$ [au]. 
The theoretical radiants of this group occupy a big part of the celestial sphere { and} their potential activities are possible all year { round}.

Among our Cyclids-SEAs we found $6$ of $13$ listed in \citet{2008MNRAS.386.2031B}, { however,} it should be noted that, { in order} to select the SEA, \citet{2008MNRAS.386.2031B} used the criteria: $a\,\epsilon\,[0.95, 1.05]$ [au], $e\,\epsilon\,[0, 0.1]$ and $i\,\epsilon\,[0, 10^\circ]$ { which}  are essentially different { to those that are} applied in our study. 

{ Additionally,}  we discovered all $5$ members of the C3 group { that were} identified by \citet{2012Icar..220.1050S} { among our Cyclids-SEAs}. We do not consider { the} Cyclids-SEAs group as a family of NEAs. We simply report that in our study all members of { the} C3 group belong to the most numerous association. 
\subsection{1994GV association }
\begin{figure}
\begin{center}
\vbox{
\includegraphics[width=0.30\textwidth]{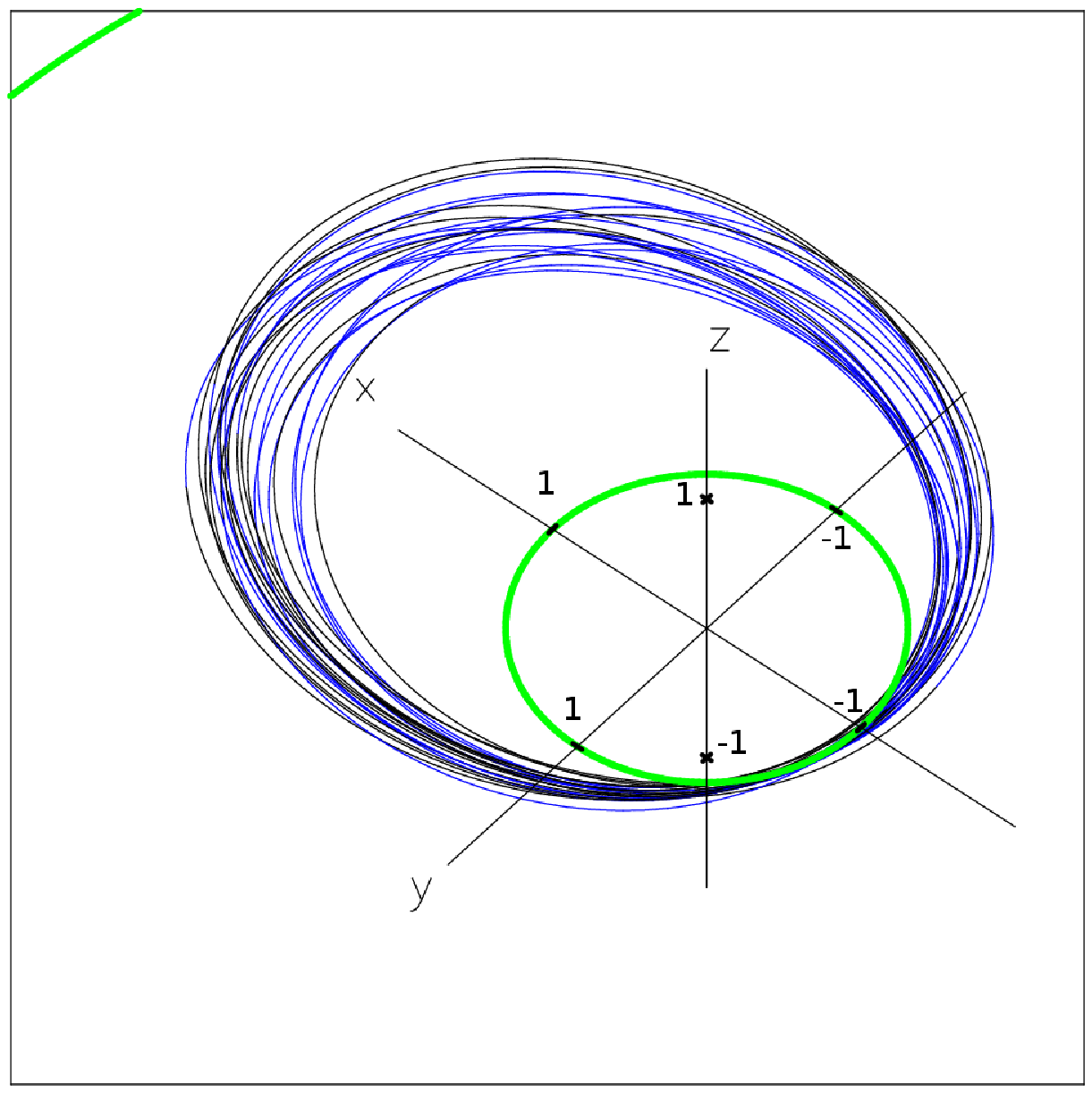}    
\\
\includegraphics[width=0.30\textwidth]{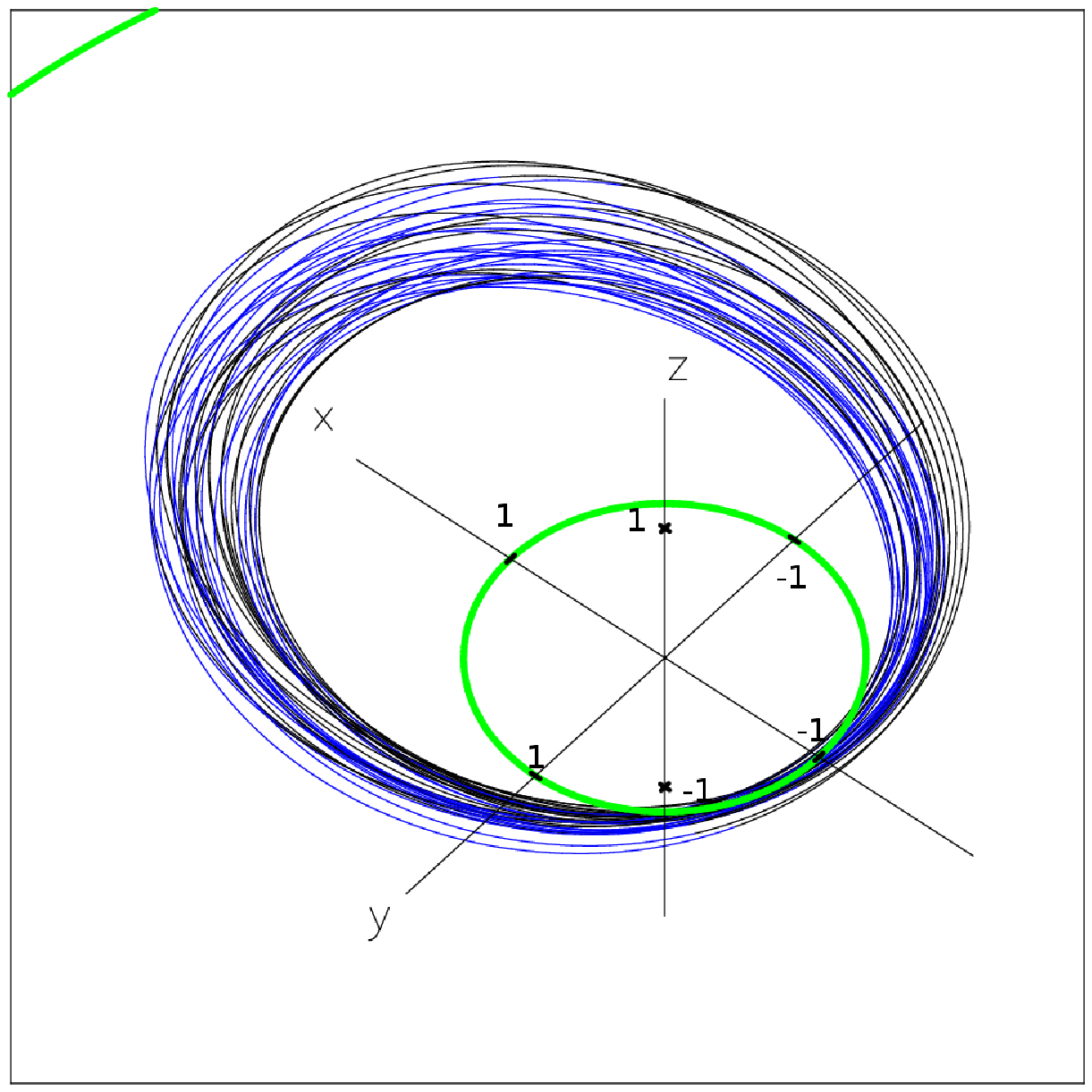} 
}
\end{center}
\caption{$58$ orbits of the 1994GV association were identified using { the} $D_H$ function and  $D_t=0.047529$. 
{ Top} --- $23$ orbits observed before AD $2012$, { bottom} --- $35$ orbits observed { between} $2012-2019$. The potential bolide activity of this group ranges from { February} 2nd to April 14th. { Earth's} orbit and { the} small arc of Jupiter's orbit are plotted in green. }
\label{fig:1994GV}
\end{figure}

This group ($D_H$/2834) was found with { the} $D_H$ and $D_t$$=$$0.047529$ { and} $58$ orbits resemble { those of} { a} meteoroid stream (see Fig.~\ref{fig:1994GV}). Using { the} $\rho_1$ metric only $4$ NEAs proved to be  members of this group. The MOIDs of the orbits amount to $0.000420$, $0.027421$, $0.156182$ [au] --- the minimal, mean and maximal, respectively. The smallest MOID { is} $0.163452$ [LD] { and} was found for 1994GV ($H$$=$$27.35^m$, size $11$ [m]), the radiant parameters of this asteroid are: $\alpha_G$$=$$69^\circ$, $\delta_G$$=$$24^\circ$ and $V_G$$=$$8.41$ [km/s], the date of possible activity is { April} 14th. 
The most massive member of this association is 2012XY6, { and} its absolute brightness and size amount to $19.09^m$, $500$ [m]. The names of the { association members are  listed} in Table~\ref{tab:1994GV}{;} $34$ of them belong to { the} {Chelyabinsk} class. 
\begin{table}
\scriptsize
\begin{center}
\caption{1994GV association. $58$ members were identified by { the} $D_H$ function and { the} single linkage cluster analysis algorithm. { The} $34$ members of { the} Chelyabinsk class are in bold.}
\begin{tabular}{l l l l ll}
\hline
    1994GV      & \sz{2007BC8   } & \sz{2007BZ48  } &     2008CF      &     2008EW5     & \sz{2008FL7   } \\ 
\sz{2008GE128 } & \sz{2009BA11  } &     2009BG81    & \sz{2009CD2   } & \sz{2009CV5   } & \sz{2009DN45  } \\ 
\sz{2009DW    } & \sz{2009FQ    } &     2009FR      & \sz{2010GL65  } & \sz{2011BE24  } & \sz{2011BV10  } \\ 
\sz{2011CZ6   } &     2011FQ16    & \sz{2012DW32  } &     2012FM      & \sz{2012FQ52  } & \sz{2012XO111 } \\ 
\sz{2012XY6   } &     2013DS9     & \sz{2013ER    } & \sz{2013GD55  } &     2014CB13    & \sz{2014DP21  } \\ 
    2014HK196   & \sz{2014HZ164 } & \sz{2015AZ43  } & \sz{2015CQ13  } &     2015CS      & \sz{2015DD54  } \\ 
\sz{2015EP    } &     2015FX35    & \sz{2015HW11  } & \sz{2016CB138 } &     2016CJ30    &     2016GD2     \\ 
    2016GF1     &     2016GL134   & \sz{2017BX6   } &     2018CF3     & \sz{2018CG3   } &     2018CH2     \\ 
\sz{2018FN3   } &     2018GN1     &     2018HD1     &     2018XQ4     & \sz{2019BD3   } & \sz{2019BH3   } \\ 
    2019CP4     & \sz{2019ET2   } &     2019GD4     &     2019GS5     &                 &   \\ 
\hline
\label{tab:1994GV}
\end{tabular}
\end{center}
\normalsize
\end{table} 
\subsection{(243147) association}
\begin{figure}
\centerline{
\hbox{
\includegraphics[width=0.30\textwidth]{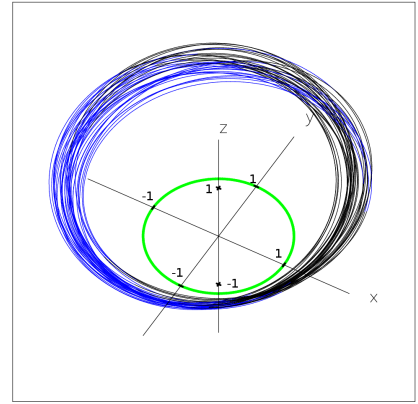}
}
}
\caption{$38$ orbits of (243147) association { were} identified with { the} $D_H$ function and $D_t=0.046288$. Possible activities of the radiants of this association range from { July} 12th until { August} 12th. { Earth's} orbit is plotted in green.}
\label{fig:243147}
\end{figure}

This group ($D_H$/1007) of $38$ NEAs was found with { the} $D_H$ and $D_t=0.046288$. The dominant body is (243147), its absolute brightness and size are $17.25^m$ and $112$ [m].  The orbits resemble { a} meteoroid stream (see Fig.~\ref{fig:243147}), { however,} they did not approach { Earth's} orbit closely. The minimal, mean and maximal MOIDs amount to $0.078342$, $0.187735$ and  $0.289566$ [au], { respectively}. The minimal MOID ($30.488$ [LD]) was found for { the asteroid} 2015PQ56  ($H$$=$$22.56^m$, diameter $D$$=$$110$ [m]). { The} corresponding radiant parameters are $\alpha_G$$=$$203^\circ$, $\delta_G$$=$$-50^\circ$ and $V_G$$=$$6.48$ [km/s], the date of possible activity of the radiant is { August} 12th. The names of the NEAs falling into this association are listed in Table \ref{tab:243147}. 
 No member of the group belongs to { the} { Chelyabinsk} class.%
\begin{table}
\scriptsize
\begin{center}
\caption{(243147) association. $38$ members were identified by { the} $D_H$ function and { the} single linkage cluster analysis algorithm.}
\begin{tabular}{l l l l ll}
\hline
243147     & 436035     & 436094     & 439889     & 2003NB     & 2006QB58   \\ 
2007PQ     & 2009MC9    & 2010RD31   & 2011NZ     & 2011OC57   & 2011PU     \\ 
2011PX     & 2012PO17   & 2013MX5    & 2013OT3    & 2014MK6    & 2014PR62   \\ 
2015PQ56   & 2015RX2    & 2016ND33   & 2016ND56   & 2016NH15   & 2016NH22   \\ 
2016NT32   & 2016NW32   & 2016PH1    & 2016RF17   & 2016TD10   & 2016TX1    \\ 
2016UX25   & 2017MB4    & 2017RF15   & 2017RT15   & 2018ME7    & 2018PX7    \\ 
2018RR3    & 2018TU2    &            &            &           &  \\ 
\hline
\label{tab:243147}
\end{tabular}
\end{center}
\normalsize
\end{table} 
\subsection{1999TV16 association }
\begin{figure}
\begin{center}
\vbox{
\includegraphics[width=0.30\textwidth]{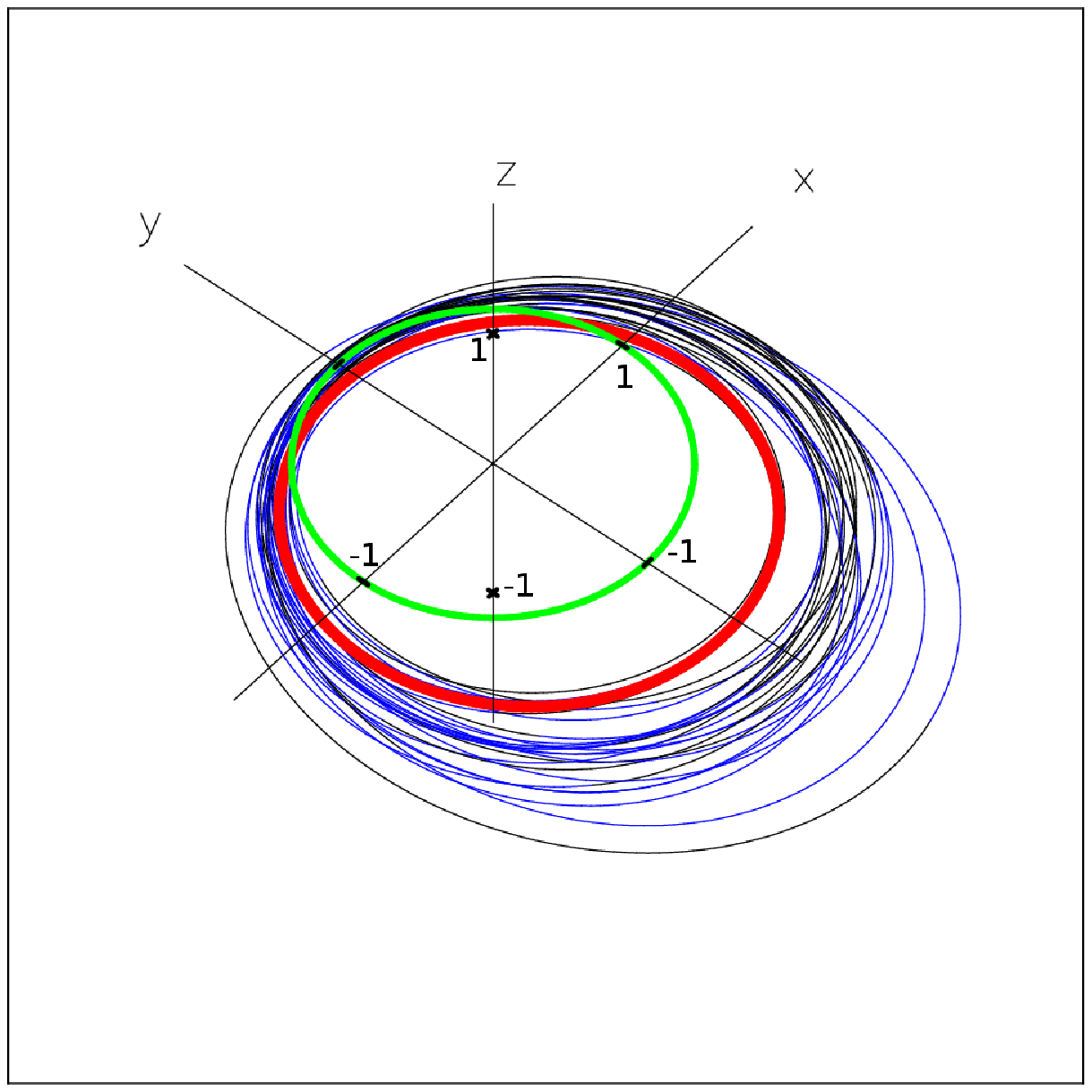}    
\\
\includegraphics[width=0.30\textwidth]{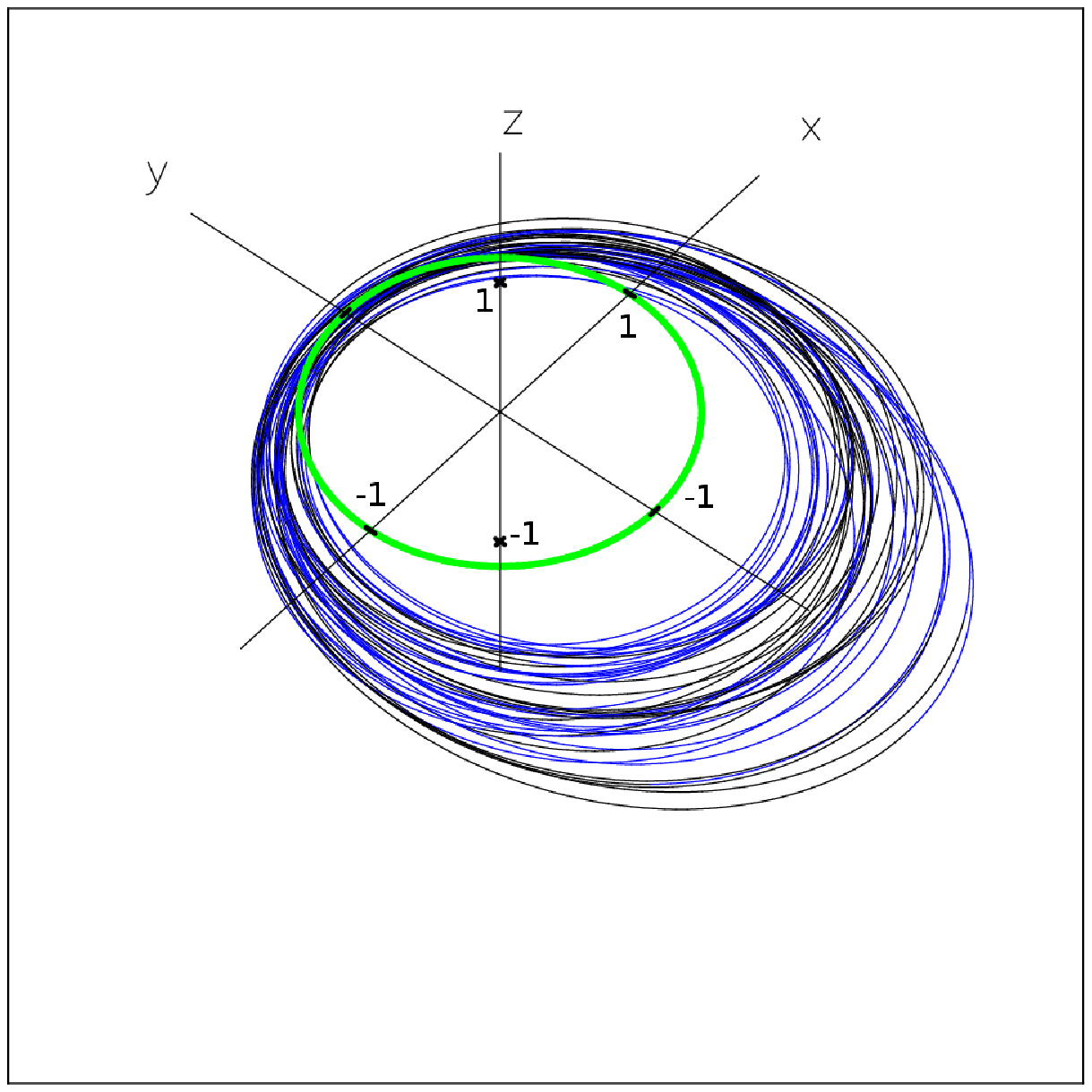} 
}
\end{center}
\caption{$69$ orbits of { the} 1999TV16 association were identified with { the} $\rho_2$ function and  $D_t=0.056485$. 
 { Top} --- $27$ orbits observed before AD $2012$, { bottom} --- $42$ orbits observed { between} $2012-2019$. The potential bolide activity of this group ranges from { October} 7th to { February} 8th. The red { colour} orbit corresponds to 2008TC3 which entered { Earth's} atmosphere on { October} 7th 2008. It was the first predicted and observed collision of the NEA with our planet. { Earth's} orbit is plotted in green.}
\label{fig:1999TV16}
\end{figure}

The group of $69$ NEAs ($\rho_2$/3041) was found using { the} $\rho_2$ distance function and the threshold $D_t=0.056485$. 1999TV16 is { a} small object { with} diameter $90$ [m] and absolute magnitude $H=23.40^m$. 
Fig.~\ref{fig:1999TV16} { presents} two plots of the { group's orbit}, one for $27$ objects discovered before $2012$ and { the other} for $42$ objects discovered since $2012$. It is easy to agree { that} the drawings represent the same asteroid stream. 
The orbits approach { Earth's} orbit quite closely, the minimal, mean and maximal MOIDs are $0.000031$, $0.02011$ and $0.11531$ [au], respectively.   
The smallest MOID ($0.012$ [LD]) corresponds to 2008TC3, the radiant parameters are $\alpha_G=348^\circ$, $\delta_G=8^\circ$ and $V_G=6.69$ [km/s], the date of possible activity of the radiant is October 7th. { The members' names are listed} in Table~\ref{tab:1999TV16}.
The asteroid 2008TC3 was a very small $4.1$ [m] size object observed during { only} 1 day in $2008$.
On October~7th, it entered { Earth's} atmosphere and exploded above the Nubian Desert in Sudan.
The impact had been predicted prior to its entry into the atmosphere as a meteor. 
A few { hundred}  meteorites, collectively named Almahata Sitta, were recovered by Peter Jenniskens, with help from students and staff of the University of Khartoum, \citep[see][]{2009DPS....41.0901J, 2009Natur.458..485J}.
The case of 2008TC3 shows that the existence of NEAs associations in { close proximity to} { Earth} may be a real threat to its inhabitants.
\begin{table}
\scriptsize
\begin{center}
\caption{1999TV16 association. $69$ members were identified by { the} $\rho_2$ function and { the} single linkage cluster analysis algorithm. The group has $24$ members meeting { the Chelyabinsk} class criteria. Their names are in bold.}
\begin{tabular}{l l l l ll}
\hline
\sz{1999TV16  } & \sz{2001AV43  } &     2003YT70    &     2005WF55    &     2006XQ4     & \sz{2006XY    } \\ 
    2006YH2     &     2007DC      &     2007TX22    &     2008TC3     & \sz{2008VA15  } &     2008WM61    \\ 
    2009BF58    & \sz{2009RH    } & \sz{2009UK14  } &     2009WK1     & \sz{2009WV7   } & \sz{2010AR1   } \\ 
    2010DL      &     2010VO21    &     2010WW8     &     2010XC      &     2010XF64    &     2010XG64    \\ 
    2011BY10    &     2012BA102   &     2012BU1     &     2012VJ38    &     2012XN134   & \sz{2013AC    } \\ 
    2013AC53    &     2013AP20    &     2013VO4     & \sz{2013VX4   } & \sz{2014AE5   } &     2014AE16    \\ 
\sz{2014CS13  } &     2014SH224   &     2014UW57    &     2014WE121   &     2015DD1     &     2015VH2     \\ 
    2015VR64    & \sz{2015VV2   } &     2015WK      &     2015XN      &     2016AN66    & \sz{2016BF1   } \\ 
    2016DK1     &     2016TA11    & \sz{2016TG55  } & \sz{2016UB107 } &     2016UN36    &     2016WZ8     \\ 
    2017AD21    &     2017BG92    &     2017ST19    &     2017VV12    & \sz{2017XR60  } & \sz{2017YD    } \\ 
\sz{2017YF8   } & \sz{2017YR3   } & \sz{2017YR4   } &     2018BD5     &     2018VX1     &     2019BO      \\ 
\sz{2019CR1   } & \sz{2019CV1   } & \sz{2019CZ2   } &                 &                 &   \\ 
\hline
\label{tab:1999TV16}
\end{tabular}
\end{center}
\normalsize
\end{table} 
\subsection{2000BH19 association}
\begin{figure}
\centerline{
\hbox{
\includegraphics[width=0.30\textwidth]{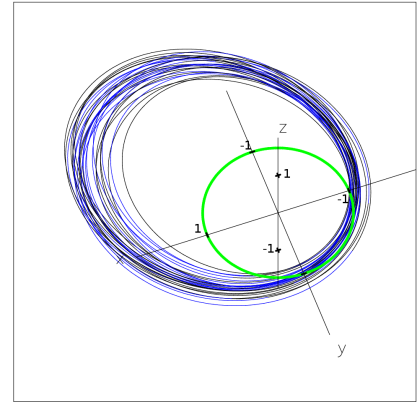} 
}
}
\caption{$30$ orbits of 2000BH19 association { were} identified using { the} $\rho_2$ and $D_H$ functions. The configuration of { the} orbits resembles a meteoroid stream. { Earth's} orbit is plotted in green.}
\label{fig:2000BH19}
\end{figure}

The group of $30$ orbits ($\rho_2$/3065) was found with { the} $D_H$ and $\rho_2$ functions, the thresholds were $0.045382$ and $0.054300$, respectively. The dominant member is 2000BH19 { with} absolute magnitude $H=19,36^m$, and diameter $400$ [m]. The orbits resemble a compact meteoroid stream, see Fig.~\ref{fig:2000BH19}. The  minimal, mean and maximal MOIDs of this group are:  $0.000280$, $0.022864$, $0.157994$ [au], { respectively}. The smallest MOID, $0.108968$ [LD], { was} found for 2018FL29, $H=30.07^m$. { The} theoretical radiant parameters of this object are $\alpha_G=38^\circ$, $\delta_G=15^\circ$ and $V_G=11.91$ [km/s], the date of possible activity is { November} 21st. The association members { are listed} in Table~\ref{tab:2000BH19}. $16$ members are  of { the} Chelyabinsk class.
\begin{table}
\scriptsize
\begin{center}
\caption{Association 2000BH19. $30$ members were identified by { the} $\rho_2$ function and { the} single linkage cluster analysis algorithm. $16$ members fall into { the} Chelyabinsk class.}
\begin{tabular}{l l l l ll}
\hline
    2000BH19    &     2003XV      &     2006AH4     & \sz{2006WM3   } & \sz{2009FG    } &     2010XN69    \\ 
\sz{2011WS95  } &     2011YB63    & \sz{2012VO6   } & \sz{2014AZ32  } &     2014DK23    & \sz{2014FX6   } \\ 
    2014YD15    &     2015FU344   &     2015XV384   & \sz{2016CW246 } & \sz{2016XA    } & \sz{2016XK    } \\ 
\sz{2016XV17  } & \sz{2017BN92  } &     2017WH13    &     2018BE3     &     2018FL29    & \sz{2018GV3   } \\ 
\sz{2018PU24  } & \sz{2018UQ1   } &     2018VV1     & \sz{2018VX6   } &     2018WE      & \sz{2019AB5   } \\ 
\hline
\label{tab:2000BH19}
\end{tabular}
\end{center}
\normalsize
\end{table} 
\subsection{(523606) association}
\begin{figure}
\centerline{
\hbox{
\includegraphics[width=0.30\textwidth]{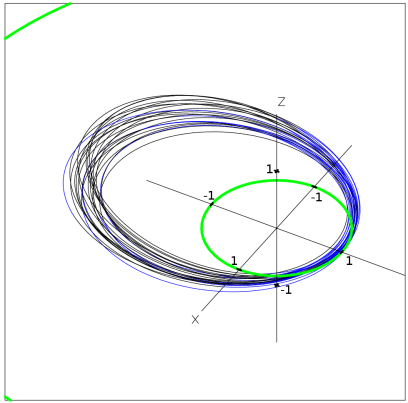}
}
}
\caption{$24$ orbits of (523606) association were identified with { the} $\rho_1$ function and  $D_t=0.049453$.  The potential bolide activity of this group ranges from October { 9th} to January 13th. { Earth's} orbit is { plotted in} green.}
\label{fig:523606}
\end{figure}

{ Eight} members of { the} (523606) association were identified { using} { the} $D_H$ function and  $D_t=0.034668$. { A} similar result was obtained with { the} $\rho_2$ and $D_t=0.044688$ { where} $11$ objects were found. However, using { the} $\rho_1$ 
and $D_t=0.049453$, $24$ NEAs were classified as an association ($\rho_1$/2710). { The asteroid} (523606) is the biggest and brightest object in this group, $H$$=$$20.23^m$, $D=300$ [m]. The spatial distribution of { this cluster's orbits} resembles  a meteoroid stream, see Fig.~\ref{fig:523606}. 
%
\begin{table}
\scriptsize
\begin{center}
\caption{(523606) association. $24$ members were identified by { the} $\rho_1$ function and { the} single linkage cluster analysis algorithm. $14$ NEAs in { bold} belong to { the} { Chelyabinsk} class.}
\begin{tabular}{l l l l ll}
\hline
\sz{523606    } & \sz{1993TZ    } & \sz{2003SS84  } & \sz{2006BN26  } & \sz{2007UF6   } & \sz{2007UU3   } \\ 
\sz{2007VD184 } & \sz{2009UL28  } &     2009UW87    &     2011WQ4     & \sz{2012CU    } & \sz{2013BD74  } \\ 
\sz{2013UE1   } & \sz{2014SC324 } &     2015BP513   & \sz{2016XA21  } &     2017AP4     &     2017TH5     \\ 
\sz{2017TZ3   } &     2017UU2     &     2018BQ5     &     2018UA      &     2018VD5     &     2018VR1     \\ 
\hline
\label{tab:523606}
\end{tabular}
\end{center}
\normalsize
\end{table} 
All orbits approach { Earth's} orbit at a distance { that is less} than $0.05$ [au] (PHA limit); the minimal,  mean and maximal MOIDs are: $0.000049$, $0.019069$,  $0.026283$ [au], { respectively} . The smallest MOID { was} found for 2018UA, $H=30.15^m$. { The} theoretical radiant parameters of this object are $\alpha_G$$=$$354^\circ$, $\delta_G$$=$$18^\circ$ and $V_G$$=$$11.92$ [km/s], the date of possible activity is { October} 20th. In [LD] unit 2018UA MOID$=$$0.019069$, which equals $1.15$ of { Earth's} radius.   

{ Table}~\ref{tab:523606} { lists} the names of $24$ members of the { group;} $14$ of them belong to { the} { Chelyabinsk} class.  
\subsection{2000SE8 association}
\begin{figure}
\centerline{
\hbox{
\includegraphics[width=0.30\textwidth]{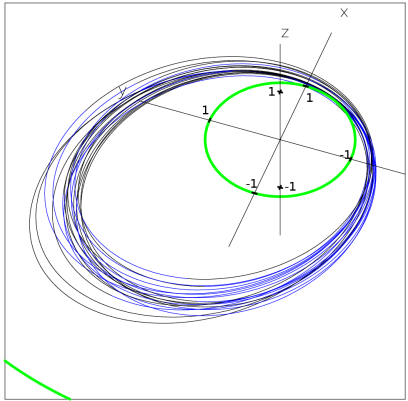} 
}
}
\caption{$20$ orbits of 2000SE8 association { were} identified using { the} $\rho_1$ function. The configuration of orbits resembles a meteoroid stream. { Earth's} and Jupiter's orbits are plotted in green. }
\label{fig:2000SE8}
\end{figure}

Twenty orbits were found with { the} $\rho_1$ function and $D_t=0.048248$ ($\rho_1$/3137). The NEAs of this group are rather small objects; one of the dominant { members} is 2000SE8  { with} absolute magnitude $H$$=$$22.9^m$, and size $60$ [m].  
The orbits resemble a meteoroid stream, see Fig.~\ref{fig:2000SE8}. The  minimal, mean and maximal MOIDs of this group are: $0.000257$, $0.045534$,  $0.129974$ [au], { respectively}.  The smallest MOID ($0.10017$ [LD]) { was} found for { the asteroid} 2015SG, $H$$=$$26.43^m$. { The} theoretical radiant parameters of this object are $\alpha_G$$=$$232^\circ$, $\delta_G$$=$$-21^\circ$ and $V_G$$=$$7.77$ [km/s], the date of possible activity is { September} 5th. The { group's}  members { are listed} in { Table}~\ref{tab:2000SE8}; half of them are { of { the} Chelyabinsk} class.
\begin{table}
\scriptsize
\begin{center}
\caption{2000SE8 association. $20$ members were identified by { the} $\rho_1$ function and a single linkage cluster analysis algorithm. { The} names of $10$ {Chelyabinsk-class} objects are in bold.}
\begin{tabular}{l l l l ll}
\hline
\sz{2000SE8   } & \sz{2007RR17  } &     2009SM1     & \sz{2010RB130 } &     2010RC31    &     2010RQ64    \\ 
    2011OE      &     2011QH50    & \sz{2014SS1   } & \sz{2015RO35  } &     2015RZ36    &     2015SG      \\ 
\sz{2016NJ39  } & \sz{2016NR55  } &     2016PG      & \sz{2016RV19  } & \sz{2017PD25  } &     2017QS32    \\ 
\sz{2017SB12  } &     2018QR      &                 &                 &                 &  \\ 
\hline
\label{tab:2000SE8}
\end{tabular}
\end{center}
\normalsize
\end{table} 
\subsection{1999YD association}
\begin{figure}
\centerline{
\hbox{
\includegraphics[width=0.30\textwidth]{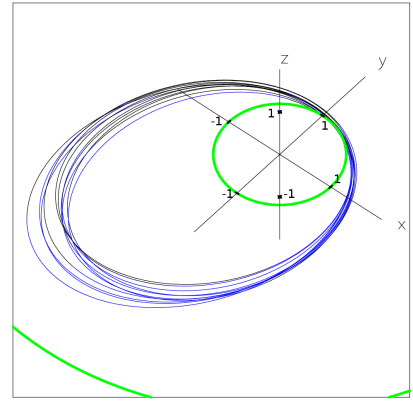} 
}
}
\caption{$11$ orbits of the 1999YD association were identified using { the} $D_H$ and $\rho_1$ { functions}. The configuration of the  orbits resembles { a} compact meteoroid stream. { Earth's orbit} and Jupiter's orbit are plotted in green.}
\label{fig:1999YD}
\end{figure}

{ We obtained the exact} same result using { the} $D_H$ and $\rho_1$ { functions;} $11$ orbits of this association ($D_H$/3058) were found with { the} thresholds $0.037268$ and $0.043102$, respectively. The largest member of this group is  1999YD, $H=21.12^m$, size $200$ [m].  The association clearly { resembles} a compact meteoroid stream, see Fig.~\ref{fig:1999YD}. The orbits approach { Earth's} orbit closely; the minimal, mean and maximal MOIDs are: $0.000342$, $0.032855$, $0.083527$ [au], { respectively}. The smallest MOID ($0.133096$ [LD]) { was} found for 2007YM, $H=26.06^m$. { The} theoretical radiant parameters of this object are $\alpha_G=351^\circ$, $\delta_G=-9^\circ$ and $V_G=8.34$ [km/s]; the date of possible activity is { November} 21st. The association members { are listed} in Table~\ref{tab:1999YD}; $6$ of them are of { the} Chelyabinsk class.
\begin{table}
\scriptsize
\begin{center}
\caption{1999YD association. $11$ members were identified by { the} $D_H$ and $\rho_1$ functions and { the} single linkage cluster analysis algorithm. $6$ NEAs { meet the} criteria of { the} {Chelyabinsk} class.}
\begin{tabular}{l l l l ll}
\hline
\sz{1999YD    } & \sz{2000WM10  } &     2003UW5     & \sz{2004YE    } &     2007YM      & \sz{2008UT5   } \\ 
\sz{2008YW32  } &     2016WV2     & \sz{2017UT7   } &     2018TS      &     2018VQ6     &  \\ 
\hline
\label{tab:1999YD}
\end{tabular}
\end{center}
\normalsize
\end{table} 
\subsection{1996GQ association}
\begin{figure}
\centerline{
\hbox{
\includegraphics[width=0.30\textwidth]{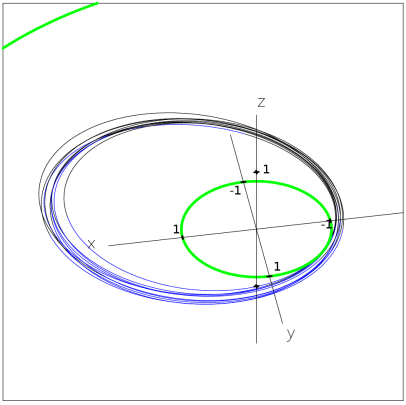} 
}
}
\caption{$10$ orbits of 1996GQ association were identified using { only} { the} $\rho_1$ function; { the} applied orbital similarity threshold was $D_t=0.043102$. { The} configuration of the orbits clearly resembles a meteoroid stream.
{ Earth's orbit} and Jupiter's orbit are plotted in green.}
\label{fig:1996GQ}
\end{figure}

The group consists of $10$ orbits { and} was found using { only} { the} $\rho_1$ function and $D_t=0.043102$  (internal code is $\rho_1$/$2870$).  The brightness of { the asteroid} 1996GQ is $23.17^m$, and its size equals $210$ [m] (with { the} assumed  albedo  of $0.024$).  
The orbits clearly resemble { those of} a meteoroid stream, see Fig.~\ref{fig:1996GQ}. The minimal, mean  and maximal MOIDs of this group are: $0.008514$, $0.062156$, $0.134925$ [au], { respectively}.  The smallest MOID ($3.313396$ [LD]) { was} found for 2009FQ10, $H$$=$$25.51^m$. { The} theoretical radiant parameters of this object are $\alpha_G$$=$$46^\circ$, $\delta_G$$=$$11^\circ$, $V_G$$=$$6.11$ [km/s] and the date of possible activity is { February} 13th. The association members { are listed} in { Table}~\ref{tab:1996GQ}, $3$ objects belong to { the} { Chelyabinsk} class.
\begin{table}
\scriptsize
\begin{center}
\caption{1996GQ association. $10$ members were identified by { the} $\rho_1$ function, $D_t=0.043102$, and { the} single linkage cluster analysis algorithm. $3$ asteroids { in bold meet the} criteria of { the} { Chelyabinsk} class.}
\begin{tabular}{l l l l ll}
\hline
\sz{1996GQ    } & \sz{1999FQ10  } &     2012FO52    &     2013DX      &     2014ES3     &     2016CY193   \\ 
    2016DM2     &     2017BL123   & \sz{2017EJ4   } &     2019CT      &                 &   \\ 
\hline
\label{tab:1996GQ}
\end{tabular}
\end{center}
\normalsize
\end{table} 
\subsection{2014FW32 association}
\begin{figure}
\centerline{
\hbox{
\includegraphics[width=0.30\textwidth]{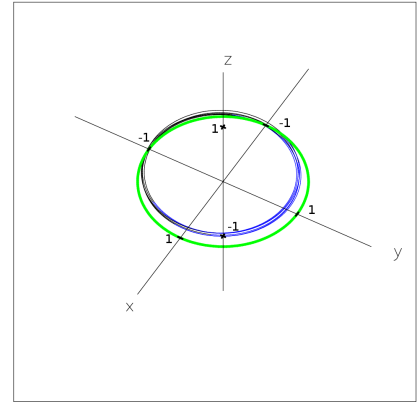} 
}
}
\caption{{ A} group of $6$ orbits of 2014FW32 association { was} identified using { the} $\rho_1$ and $\rho_2$ functions. The orbits resemble Cyclids-SEAs association and are located inside  { Earth's} orbit, the biggest ellipse marked in green.}
\label{fig:2014FW32}
\end{figure}

The group ($\rho_1$/10793) consists of $6$ orbits, it was found using { the} $\rho_1$ and $\rho_2$ functions, the thresholds were $D_t$$=$$0.037021$ and $Dt$$=$$0.036307$. The NEA 2014FW32 absolute magnitude is $26.98^m$ and its diameter is $15$ [m].  
The orbits resemble { those of}  a meteoroid stream, see Fig.~\ref{fig:2014FW32}, similar to Cyclids-SEAs association, { but} all the orbits are within { Earth's} orbit. All $6$ members belong to { the} Aten group.  The  minimal, mean and maximal MOIDs of this association are:  $0.000702$, $0.007543$, $0.015019$ [au], { respectively}.  The smallest MOID ($0.273198$ [LD]) { was} found for 2018GR4, $H$$=$$27.08^m$. { The} theoretical radiant parameters of this object are $\alpha_G$$=$$328^\circ$, $\delta_G$$=$$-25^\circ$ and $V_G$$=$$2.54$ [km/s], the date of possible activity is March 11th. The members of this association { are listed} in { Table}~\ref{tab:2014FW32}. Only one asteroid of this group belongs to { the} {Chelyabinsk} class.
\begin{table}
\scriptsize
\begin{center}
\caption{2014FW32 association. $6$ members were identified by { the} $\rho_1$ and $\rho_2$ functions. { The} single linkage cluster analysis algorithm was applied. { The} name of the { Chelyabinsk-class} asteroid is in bold.}
\begin{tabular}{l l l l ll}
\hline
    2014FW32    &     2014HN2     &     2016HF19    & \sz{2017HK1   } &     2018GE      &     2018GR4   \\ 
\hline
\label{tab:2014FW32}
\end{tabular}
\end{center}
\normalsize
\end{table} 
\section{Conclusions}
We have made three attempts at { a search that was} extensive and very rigorous { search} for grouping among $20032$ NEAs. {} The probability of accidental clustering was small, below $1$\%. { Fifteen} associations were found using { the} single linkage cluster analysis algorithm, { which is} the same { algorithm that} \citet{1990AJ....100.2030Z}  used { in their} search for asteroid  families among the main belt asteroids (MBA). Zappala et al. called this algorithm { a} hierarchical cluster analysis algorithm. In our { search} only $953$  orbits ($\sim4.7$\%) were classified as association members. 

To quantify the similarity among { the} two orbits we applied three distance functions: $D_H$, $\rho_1$ and $\rho_2$. Both $\rho$-functions proposed recently by \citet{2016MNRAS.462.2275K} worked well, { and} they { provided} results comparable with { those} obtained by { the} $D_H$ function \citep{1993Icar..106..603J}. { Thus}, in our view, the new functions do not constitute a major breakthrough in small bodies cluster analysis. However, from { the} mathematical { perspective}, they have the obvious advantage { satisfying} three axioms of a metric space. Therefore, they must be applied { in} the first { instance}.

All orbital plots of { the} identified associations clearly resemble meteoroid streams, see Figures \ref{fig:Anza23}--\ref{fig:Didymos} and \ref{fig:Cyclids}--\ref{fig:2014FW32}. However, we do not claim that our associations are of { a} genetic nature, i.e. that they originated from { a} common parent body. No cluster analysis can prove { this}. We only { claim that,}  despite their origin, { the} identified associations { do} exist. Currently, we do not know how many members a single association contains. { It could be} $70$ members, but perhaps it contains $700$, { for example}. 
We showed that the { number} of members identified in { an} association increases { over}  time, as { was the case for} Anza, Zephyr, Toutatis, Itokawa and other large groups. 
This { is an} important { fact} if the probability of the collision between { Earth} and the NEAs is to be calculated. { As aforementioned,} the threat to { Earth} and its inhabitants is very serious. Among the members of the identified associations we found many members, on average $34$ \%, { which}  are larger than the Chelyabinsk object and which  approach { Earth's}  orbit { at a distance smaller} than $0.05$ [au] (PHA limit). Two asteroids { in these}  associations approached { Earth's} orbit at extremely close distances { which are shorter} than { Earth's} radius: 2017FU102 (a member of { the} Itokawa association) and 2008TC3 (a member of 1999TV16 association). The latter object entered { Earth's} atmosphere on October 7 2008 and exploded at $\sim37$ [km]  above the Nubian Desert in { Sudan} \citep{2009Natur.458..485J}.

The common origin of { the} members of { the} identified groups is an open question. They could { have emerged} as a result of collisions  (e.g. Itokawa and Toutatis associations) or { as a} result of migration processes { that were} induced by some dynamical resonances. Further investigations are needed on the number and the nature of the NEAs associations. 
\vspace{-0.5cm}
\section*{Acknowledgements}
The author acknowledges the anonymous referee for the comments and suggestions.
T.J. Jopek was supported by the National Science { Centre} in Poland (project No 2016/21/B/ST9/01479).
This research has made use of NASA's Astrophysics Data System Bibliographic Services. 
\vspace{-0.5cm}
%

\bibliographystyle{mnras}
\bibliography{nea_grupy_1.bib}\label{refs} 


\bsp	
\label{lastpage}
\end{document}